\documentclass[twocolumn]{aastex631}

\usepackage{multirow}
\usepackage{amsmath}
\usepackage{amssymb}
\usepackage{comment}
\usepackage{bm}

\def\Rmrgbh{R_{\bullet\bullet}}

\begin{document}

\title{Multiband gravitational wave observations of eccentric escaping binary black holes from globular clusters}

\author[0000-0002-6154-4381]{Yuetong Zhao}
\altaffiliation{ytzhao@bnu.edu.cn}
\affiliation{School of physics and astronomy, Beijing Normal University, 19 Xinjiekouwai St, Beijing 100875,China \\}

\author[0000-0001-9688-3458]{Abbas Askar}
\altaffiliation{askar@camk.edu.pl}
\affiliation{Nicolaus Copernicus Astronomical Center, Polish Academy of Sciences, ul. Bartycka 18, PL-00-716 Warsaw, Poland\\}

\author{Youjun Lu}
\altaffiliation{luyj@nao.cas.cn}
\affiliation{CAS Key Laboratory for Computational Astrophysics, National Astronomical Observatories, Chinese Academy of Sciences,  20A Datun Road, Beijing 100101, China\\}
\affiliation{School of Astronomy and Space Sciences, University of Chinese Academy of Sciences, 19A Yuquan Road, Beijing 100049, China\\}

\author{Zhoujian Cao}
\altaffiliation{zjcao@bnu.edu.cn}
\affiliation{School of physics and astronomy, Beijing Normal University, 19 Xinjiekouwai St, Beijing 100875,China \\}

\author{Mirek Giersz}
\affiliation{Nicolaus Copernicus Astronomical Center, Polish Academy of Sciences, ul. Bartycka 18, PL-00-716 Warsaw, Poland\\}

\author{Grzegorz Wiktorowicz}
\affiliation{Nicolaus Copernicus Astronomical Center, Polish Academy of Sciences, ul. Bartycka 18, PL-00-716 Warsaw, Poland\\}

\author[0000-0003-0596-0919]{Arkadiusz Hypki}
\affiliation{Faculty of Mathematics and Computer Science, A. Mickiewicz University, Uniwersytetu Pozna\'nskiego 4, 61-614 Pozna\'n, Poland\\}

\author{Lucas Hellstrom}
\affiliation{Nicolaus Copernicus Astronomical Center, Polish Academy of Sciences, ul. Bartycka 18, PL-00-716 Warsaw, Poland\\}

\author{Sohaib Ali}
\affiliation{Nicolaus Copernicus University, Jurija Gagarina 11, 87-100 Toruń, Poland\\}

\author{Wei-Tou Ni}
\affiliation{International Centre for Theoretical Physics Asia-Pacific, University of Chinese Academy of Sciences, Beijing 100190, China\\}
\affiliation{Innovation Academy of Precision Measurement Science and Technology (APM), Wuhan Institute of Physics and Mathematics, Chinese Academy of Sciences, Wuhan 430071, China}




\begin{abstract}

Stellar-mass binary black holes (sBBHs) formed in globular clusters (GCs) are promising sources for multiband gravitational wave (GW) observations, particularly with low- and middle-frequency detectors. These sBBHs can retain detectable eccentricities when they enter the sensitivity bands of low-frequency GW observatories. We study multiband GW observations of eccentric sBBHs that escape from GC models simulated with the MOCCA code, focusing on how low- and middle-frequency detectors can constrain their eccentricities and other parameters. Using Monte Carlo simulations, we generate ten realizations of cosmic sBBHs by combining the MOCCA sample with a cosmological model for GC formation and evolution. We then assess their detectability and the precision of parameter estimation. Our results show that LISA, Taiji, the LISA-Taiji network (LT), and AMIGO could detect $0.8\pm0.7$, $11.6\pm2.0$, $15.4\pm2.7$, and $7.9\pm1.3$ escaping sBBHs, respectively, over four years, while LT-AMIGO could detect $20.6\pm3.0$ multiband sBBHs in the same period. LT and AMIGO can measure initial eccentricities with relative errors of approximately $10^{-6}$–$2\times10^{-4}$ and $10^{-3}$–$0.7$, respectively. Joint LT-AMIGO observations have a similar ability to estimate eccentricities as LT alone.

\end{abstract}


\section{Introduction} \label{sec:intro}

More than 150 stellar-mass binary black hole (sBBH) merger events have been reported by the LVK[LIGO  (Laser Interferometer Gravitational-Wave Observatory), Virgo, and KAGRA] collaborations following their third observing run, as listed in the GWTC-4.0 catalog(\citep{gwtc4}).
Before entering the high-frequency band ($\sim 10$–$300$\,Hz) of ground-based gravitational wave detectors, sBBHs at their inspiral stages emit gravitational waves (GWs) at lower frequencies ($\sim 10^{-4}$–$10$\,Hz), which may be detected by space-based milli-Hertz GW detectors (see \citealt{Sesana2016}), such as the Laser Interferometer Space Antenna (LISA; \url{https://lisa.nasa.gov}), Taiji \citep{SSPMAHuang2017}, and TianQin \citep{Tianqin2016}. In addition, deci-Hertz GW detectors—including the Astrodynamical Middle-frequency Interferometer GW Observatory (AMIGO; \citealt{Ni2018}), Big Bang Observer (BBO; \citealt{bbo2006}), DECi-hertz Interferometer Gravitational Wave Observatory (DECIGO; \citealt{decigo}), B-DECIGO \citep{bdecigo}, and lunar-based GW detectors such as the Lunar Gravitational-Wave Antenna (LGWA; \citealt{2021Harms}) and the Gravitational-wave Lunar Observatory for Cosmology (GLOC; \citealt{gloc})—will extend observational capabilities to middle frequencies. Therefore, future multiband observations of sBBH mergers will be possible, with events detected first by low-frequency, then middle-frequency, and finally high-frequency detectors, such as the Cosmic Explorer (CE; \citealt{3rdGWD}) and the Einstein Telescope (ET; \citealt{ETstudy}).

The origin of sBBHs detected by GW detectors remains a topic of active debate in the literature \citep{2016ApJ...830L..18B,2022Chenzw,2024Nicholas}. Four main formation mechanisms have been proposed for merging sBBHs. The first is the evolution of massive binary stars in the galactic field \citep{Cris2002, Cris2016, Mandel2016,giacobbo2018,spera2019,Son2020,chris2020,mandel2022,costa2023,Olejak2024}. sBBHs formed via this channel are expected to have relatively low component masses with nearly circular orbits. The second is formation in dense stellar environments, such as globular clusters (GCs), through dynamical interactions \citep[e.g.,][]{SPZ2000,Benacquista2002, Downing2010,Downing2011,Rodriguez2016, 2016ApJ...830L..18B, 2017Askar,park2017,Samsing2018, D'Orazio2018,Hong2018,rastello2019,kremer2019,diCarlo2020,sedda2021,Mapelli2022,banerjee2021,Torniamenti2022,antonini2023}. Within this channel, sBBHs can be further categorized as either escaping or in-cluster sBBHs \citep{2017Askar,2018Rodriguez,Anagnostou2020,2024arXiv241003832B}. These binaries typically have larger masses and significant eccentricities during the inspiral phase at low GW frequencies, compared to those formed via the evolution of massive binary stars. The third formation mechanism is the AGN channel, in which sBBHs are born in the accretion disks of active galactic nuclei (AGN) or near massive black holes \citep[e.g.,][]{Antonini2012, Stephan2016, Bartos2017, AGNecc2018ApJ, Samsing2022Nat}. Binaries formed in this channel may also have high masses and large orbital eccentricities due to the gas-rich environment. The fourth scenario involves the dynamical interactions of primordial black holes \citep{2017PhRvD..96l3523A, 2018ApJ...854...41K, 2018ApJ...864...61C}. Each formation channel produces sBBHs with distinct merger-rate evolution and parameter distributions (e.g., mass and spin) \citep{2018Rodriguez, 2023Adamcewicz}. In particular, the eccentricity distribution of sBBHs is a promising probe of the relative contribution from each formation mechanism. 

Multiband GW observations of sBBHs can provide rich information, both by probing the systems at different evolutionary stages and by improving the accuracy of parameter estimation \citep{Sesana2016, 2017LISA, Gerosa2019, LiuShao2020, ChenLu2021,Izumi2021,2023LRR....26....2A,ZhaoLu2023}. For eccentric sBBHs, low- and middle-frequency GW detectors are especially important, as the eccentricity decreases rapidly due to GW emission as binaries evolve toward higher frequencies. Previous studies have investigated the observability of eccentric sBBHs in the milli-Hertz band \citep[e.g.,][]{Nishizawa2017,2019XiaoF,Randall2022}. \citet{2021Holgado} explored the prospects for eccentricity estimation for GW190521-like binaries in the deci-Hertz band. 

In this paper, we study the multiband observations of eccentric sBBHs that formed and escaped from 268 GC models simulated with the MOCCA (MOnte Carlo Cluster simulAtor) code, focusing on the detectability and eccentricity constraints for sBBHs in the low-, middle-, and joint-frequency bands. The structure of this paper is as follows. In Section~\ref{sec:sample}, we describe the mock samples of escaping sBBHs from the GC models. Section~\ref{sec:snr} presents our method for estimating the signal-to-noise (S/N) ratio for the mock eccentric sBBHs, and provides estimates of the expected numbers and parameter distributions of “detectable”\footnote{Here, “detectable” refers to those sBBHs with S/N exceeding the specific threshold adopted for each GW detector or network. For simplicity, we omit quotation marks hereafter.} mock sBBHs (individually and jointly) for various GW detectors and networks across frequency bands. In Section~\ref{sec:PE}, we present the measurement precision for the physical and geometric parameters of the detectable sBBHs, as obtained by single-band GW detectors or by joint multiband observations, using the Fisher information matrix (FIM) method. We provide discussion and conclusions in Section~\ref{sec:dis_con}.

Throughout this work, we adopt the standard $\Lambda$CDM cosmological model, with $H_0 = 67.9\,{\rm km\,s^{-1}\,Mpc^{-1}}$, $\Omega_{\rm m} = 0.306$, $\Omega_{\rm k} = 0$, and $\Omega_{\Lambda} = 0.694$ \citep{Planck2016}.

\section{Mock Sample of BBHs Originating from GCs} \label{sec:sample}

Dynamically formed sBBHs from GC can be divided into different types, including in-cluster sBBHs and ejected sBBHs \citep{2017Askar}. The orbital evolution of those in-cluster sBBHs will be influenced by the cluster environment and they will eventually merge within the cluster. In this paper, we only consider those sBBHs ejected from the cluster and evolve in isolatation since this subgroup accounts for the largest fraction of all sBBHs originating from GCs. We specifically use MOCCA (MOnte Carlo Cluster simulAtor) simulation results for sBBHs escaping from GCs with a range of initial conditions, and combine these GC models with a cosmic formation and evolution model of GCs to generate a more realistic mock sample of GC-origin sBBHs.

\subsection{MOCCA code and simulated GC models} \label{subsec:mocca-code}

This work utilizes results from a large and diverse set of 268 star cluster models simulated with the \textsc{MOCCA} code \citep{Hypki2013,Giersz2013}. A subset of these models have been previously described in \citet{Giersz2025,Giersz2025b} and \citet{Wiktorowicz2025}, while many others are presented here for the first time. The selected models provide a reasonable representation of present-day GCs as well as clusters that may have dissolved and are no longer observable. 

The \textsc{MOCCA} code is based on the Monte Carlo method \citep{1971Ap&SS..13..284H,Henon1971,1982AcA....32...63S,1986Stodolkiewicz,2001MNRAS.324..218G,2003MNRAS.343..781G}, which is used to model relaxation processes and track the long-term evolution of spherically symmetric star clusters. In addition to two-body relaxation, \textsc{MOCCA} employs the \textsc{fewbody} code \citep{2004MNRAS.352....1F,2012ascl.soft08011F}, a direct \textit{N}-body integrator optimized for small-\textit{N} gravitational dynamics and scattering experiments, to simulate the outcomes of close binary-single and binary-binary encounters, without including post-Newtonian (PN) corrections. As a consequence, highly eccentric BBH mergers that may occur during resonant few-body interactions are not captured in these simulations \citep{samsing2018b,2018Rodriguez,Samsing2018}.

The gravitational potential of the GCs' host galaxy in \textsc{MOCCA} is approximated using a simplified point-mass model, with a total mass equal to the enclosed Galactic mass at the cluster's Galactocentric distance. The code also incorporates an improved approach for handling escaping objects \citep{Fukushige2000}. Binary and stellar evolution in \textsc{MOCCA} are modeled using the prescriptions from the \textsc{sse} \citep{Hurley2000} and \textsc{bse} \citep{Hurley2002} population synthesis codes, along with numerous recent improvements to these prescriptions \citep{Belloni2018,Banerjee2020,Kamlah2022}.

\subsubsection{Initial conditions for the simulated GC models}\label{sec:initial-condition}

\begin{figure}
    \centering
    \includegraphics[width=\columnwidth]{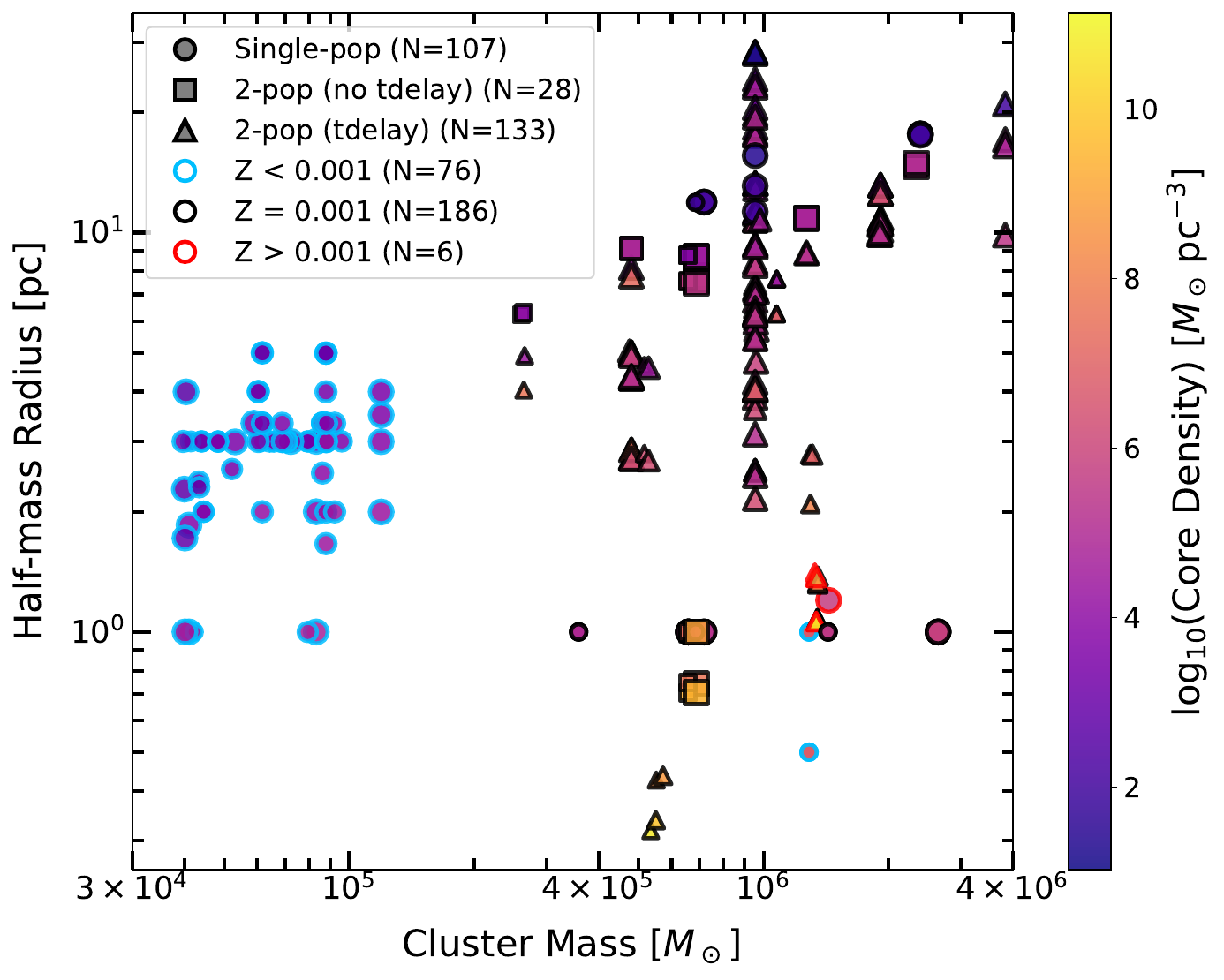}
    \includegraphics[width=\columnwidth]{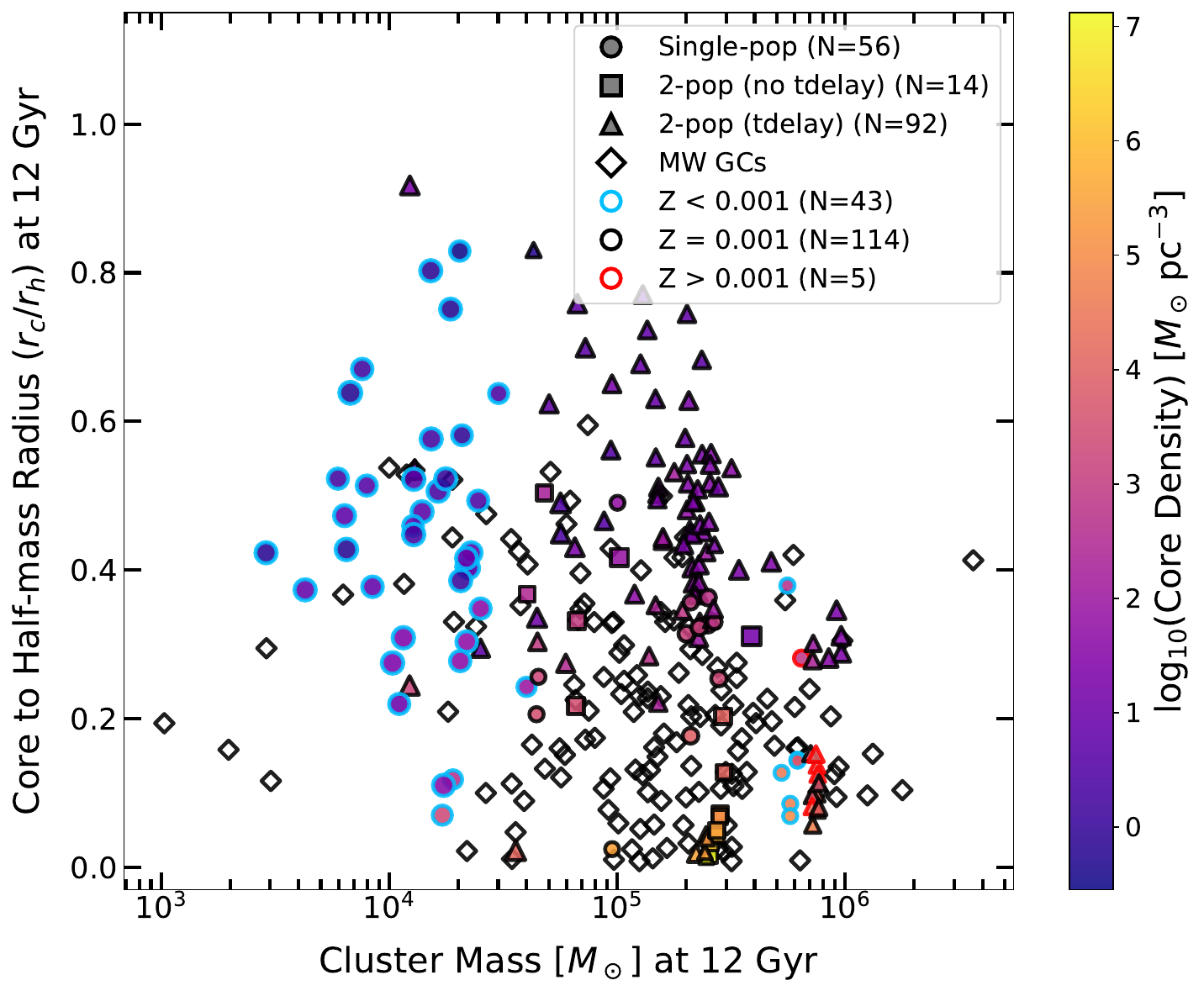}
    \caption{
        \textbf{Top:} Distribution of all MOCCA {\textbf{{GC}}} models analyzed in this work, shown in the plane of initial cluster mass versus half-mass radius. Symbol shape shows model type: single population (circle), two populations without time delay (square), and two populations with time delay (triangle). The color of each symbol represents the initial core density (see color bar), while the edge color denotes metallicity. Point size scales with the initial binary fraction, from 7\% (smallest) to 95\% (largest). \textbf{Bottom:} Core-to-half-mass radius versus cluster mass at 12 Gyr for the 165 MOCCA models that survived to this age; clusters that disrupted earlier are not shown. For comparison, black diamonds indicate observed Milky Way GCs from the \citet{Baumgardt2018} catalogue.
    }
    \label{fig:mocca-models}
\end{figure}

The 268 models used for this study span a wide range of initial conditions. The initial number of objects (singles plus binaries, with binaries counted as one object) covers $3.4\times10^4$ to $3.2\times10^6$, with corresponding total cluster masses from $3.97\times10^4\,M_\odot$ to $3.83\times10^6\,M_\odot$. A summary of the key model parameters is presented in Table~\ref{tab:model-params}. The suite includes both ``low-$N$'' models ($N_\mathrm{ini}<2\times10^5$; 70 models, $3.4\times10^4$\,--\,$1.9\times10^5$) and ``large-$N$'' models ($N_\mathrm{ini}\geq2\times10^5$; 198 models, $2.0\times10^5$\,--\,$3.2\times10^6$). The initial half-mass radii ($r_{h,\mathrm{ini}}$) extend from $0.50$ to $28.3$\,pc, core radii from $0.011$ to $14.6$\,pc, and central densities from $31.6$ to $3.85\times10^9\,M_\odot\,\mathrm{pc}^{-3}$. All clusters are initialized as King models with a variety of concentration parameters. Tidal radii range from $10.4$ to $156.5$\,pc, and initial Galactocentric radii from $0.98$ to $16.04$\,kpc (median $2.25$\,kpc). 

The initial binary fraction, defined as the ratio of initial number of binaries to total objects, ranges from $0.07$ up to $0.95$, with values of $0.10$, $0.30$, $0.35$, $0.40$, $0.50$, and $0.95$ (median $0.50$). These are grouped as low ($<$0.20; 16 models), moderate (0.20--0.50; 58 models), and high ($\geq0.50$; 194 models) binary fraction bins. Initial stellar metallicities include $Z=6.5\times10^{-5}$ (51 models), $1.0\times10^{-4}$ (19), $[2$--$7.5]\times10^{-4}$ (12), $0.001$ (186), and $[0.002$--$0.0033]$ (8 models). The median metallicity of these models is compareable to the median metallicity of GCs' observed in our Galaxy \citep[][updated 2010]{Harris1996}.

The models include 107 single-population clusters and 161 two-population clusters. The inclusion of two-population models is motivated by extensive observational evidence that most Galactic GCs host multiple stellar populations, distinguished primarily by differences in light-element abundances of their stars \citep{Bastian2018,Milone2022}. In two-population models, the first population (1P) is generally more numerous and is typically less concentrated and tidall-filling, while the second population (2P) is more centrally concentrated and sometimes introduced after a time delay. Each population is initialized as a separate King model with its own concentration parameter \citep{Hypki2022,Hypki2025,Wiktorowicz2025}. In 133 of the two-population models, the 2P is added with a time delay and is initially more centrally concentrated; see \citet{Giersz2025} for details. The remaining 28 two-population models have both populations initialized from the very beginning at t=0 Myr. The distribution of initial cluster masses and half-mass radii for the MOCCA models used in this study are shown in Figure~\ref{fig:mocca-models}, with symbol size, edge color, and shape representing the initial binary fraction, metallicity, and single or multiple population nature of each cluster, respectively. The lower panel of Figure~\ref{fig:mocca-models} shows the core-to-half-mass radius versus cluster mass at 12 Gyr for the 165 out of 268 analyzed MOCCA models that survive to this age. While these models are not specifically tailored to reproduce the present-day Milky Way GC population (shown as black unfilled diamonds), they are broadly representative of the regions of parameter space occupied by observed Galactic GCs.

Stellar masses are drawn from a \citet{Kroupa2001} initial mass function (IMF), covering $0.08$--$150\,M_\odot$ for 1P and single-population clusters, and $0.08$--$20\,M_\odot$ for 2P stars in two-population clusters. For most GC models with low initial binary fraction ($<40\%$), the initial binary parameters for systems with component masses below $\rm 5 \, M_{\odot}$ followed a uniform distribution in $\rm \log(a)$, extending up to 100 AU, with random pairing of components and a thermal eccentricity distribution. In models with a larger initial binary fraction, the initial binary period, eccentricity, and mass ratio distributions for stars less massive than $\rm 5 \, M_{\odot}$ were based on the \citet{Kroupa1995,Belloni2017} prescriptions. For all binaries, the minimum initial orbital separation was chosen to ensure that the pericenter distance was larger than the sum of the radii of the two stellar components. For all models regardless of the initial binary fraction, binaries with component masses exceeding $\rm 5 \, M_{\odot}$ followed the distributions provided by \citet{Sana2012}.

\begin{table}[htb]
    \centering
    \small
    \caption{Key initial parameters for the 268 MOCCA cluster models analyzed in this work.}
    \begin{tabular}{ll}
        \hline
        Parameter & Range / Values \\
        \hline
        Number of objects ($N_\mathrm{ini}$) & $3.4\times10^4$\,--\,$3.2\times10^6$ \\
        Total mass ($M_\mathrm{ini}$)\,[$M_\odot$] & $3.97\times10^4$\,--\,$3.83\times10^6$ \\
        Half-mass radius ($r_{h,\mathrm{ini}}$)\,[pc] & $0.50$\,--\,$28.3$ \\
        Core radius ($r_{c,\mathrm{ini}}$)\,[pc] & $0.011$\,--\,$14.6$ \\
        Central density ($\rho_{c,\mathrm{ini}}$)\,[$M_\odot$\,pc$^{-3}$] & $31.6$\,--\,$3.85\times10^9$ \\
        Tidal radius ($r_{\mathrm{tid},\mathrm{ini}}$)\,[pc] & $10.4$\,--\,$156.5$ \\
        Galactocentric radius ($R_{\mathrm{GC}}$)\,[kpc] & $0.98$\,--\,$16.04$ (median 2.25) \\
        Binary fraction ($f_{b,\mathrm{ini}}$) & $0.07$\,--\,$0.95$ \\
         & Low ($<$0.20): 16 models \\
         & Moderate (0.20--0.50):\\
         & 58 models \\
         & High ($\geq$0.50): 194 models \\
        Metallicity ($Z$) & $6.5\!\times\!10^{-5}$: 51 models \\
                  & $1.0\!\times\!10^{-4}$: 19 models \\
                  & $[2$--$7.5]\!\times\!10^{-4}$: 12 models \\
                  & $0.001$: 186 models \\
                  & $[0.002$--$0.0033]$: 6 models \\
        Two-population models & 161 / 268 \\
        \quad \hspace{0.5em}with delayed 2P addition & 133 \\
        \quad \hspace{0.5em}w/o delayed 2P addition & 28 \\
        Single-population models & 107 / 268 \\
        Low-$N$ models ($N_\mathrm{ini}<2\times10^5$) & 70; $3.4\times10^4$\,--\,$1.9\times10^5$ \\
        High-$N$ models ($N_\mathrm{ini}\geq2\times10^5$) & 198; $2.0\times10^5$\,--\,$3.2\times10^6$ \\
        \hline
    \end{tabular}
    \label{tab:model-params}
\end{table}

\subsubsection{Important prescriptions for BH and BBH production}\label{subsec:mocca-bh-prescriptions}

The simulated GC models implement several key prescriptions for massive star evolution and BH formation. Wind mass-loss rates for massive stars are metallicity dependent and modeled following \citet{vink2001,Belczynski2010}. BH and neutron star (NS) masses are determined using the `rapid' supernova model from \citet{Fryer2012}, which allows for the formation of BHs within the $ \rm 3-5 \, M_{\odot}$ first mass gap. The maximum mass of a single BH that can form through the stellar evolution of single stars in these models is $\rm 44.5 \, M_{\odot}$, which is depended on the metallicity.  

Natal kicks for NSs and BHs follow a Maxwellian distribution with $\rm \sigma = 265 \, \mathrm{km \, s^{-1}}$ \citep{Hobbsetal2005}. For BHs, these kicks are reduced according to the fallback factor, as described by \citet{Cris2002} and \citet{Fryer2012}. Pair-instability supernovae (PISNe) and pulsational pair-instability supernovae (PPISNe) are accounted for using the prescriptions of \citet{Belczynski2016}, which result in either the complete or partial disruption of BH progenitors with evolved core masses within specific ranges. Binary interactions are modeled with prescriptions for mass transfer and common-envelope evolution (with BSE common envelope parameters $\rm \alpha = 0.5$ and $\rm \lambda = 0.0$).

The birth spins of BHs are assumed to be small \citep{fullerma2019} and are drawn from a flat distribution between 0 and 0.1. MOCCA implements GW recoil kicks by calculating the kick velocity imparted when two BHs merge, using the formulae of \citet{Campanelli2007} and \citet{Baker2008}, which are fitted to numerical relativity simulations. The recoil magnitude depends on the mass ratio of the merging BHs as well as the magnitude and direction of their spins. In these models, the spin orientations of BH merger components are assumed to be distributed isotropically \citep[see][for details]{Morawski2019}. The assumed low birth spins of BHs in the population allow for a non-negligible retention fraction of second-generation (2G) BHs, since the typical recoil kicks can be small enough for merged BHs to remain bound in many clusters. However, the merger product is assigned a larger spin \citep{berti2007,hofmann2016}, so the retention probability of subsequent 2G BHs decreases significantly \citep{gerosa2017,Rodriguez2019}. Nonetheless, some models produce hierarchical mergers involving two 2G BHs, and such binaries can be dynamically ejected from the cluster in close binary-single or binary-binary encounters. These systems, if ejected and subsequently merging in the field, contribute to the highest chirp mass BBH escapers in our sample.

For a subset of 22 initially dense models, intermediate-mass BHs (IMBHs; $10^{2}$--$10^{4}\,M_\odot$) can form through the growth of a very massive star ($\gtrsim1000\,M_\odot$) via runaway stellar collisions early in the cluster's evolution, typically for models with $\rho_{c} \gtrsim 10^7\,M_\odot\,\mathrm{pc}^{-3}$ \citep[see][and references therein]{Giersz2015,Askar2023,vergara2025}. These massive stars subsequently interact with stellar-mass BHs, and 25\% of the stellar mass is assumed to be accreted onto the BH, resulting in the formation of an IMBH, which can continue to grow through mergers with stars and compact remnants.
This assumed accretion fraction may be on the conservative side, as tidal disruption studies typically find that roughly half of the stellar debris remains bound to the BH, a portion of which may subsequently be accreted, with even higher accretion efficiencies possible in more head-on collisions \citep{Baumgarte2025}. However, the true accretion efficiency remains uncertain and depends on encounter geometry, stellar structure, outflows, magnetic fields, and BH spin \citep{Kremer2022,Vynatheya2024}.
Similar to previous works, we find that clusters that form IMBHs typically yield fewer merging stellar-mass BBHs \citep{Hong2020}.  

\subsection{Mock Sample of Escaping sBBHs from MOCCA GC Models} \label{subsec:mock_sample}

\begin{figure}
\includegraphics[width=\columnwidth]{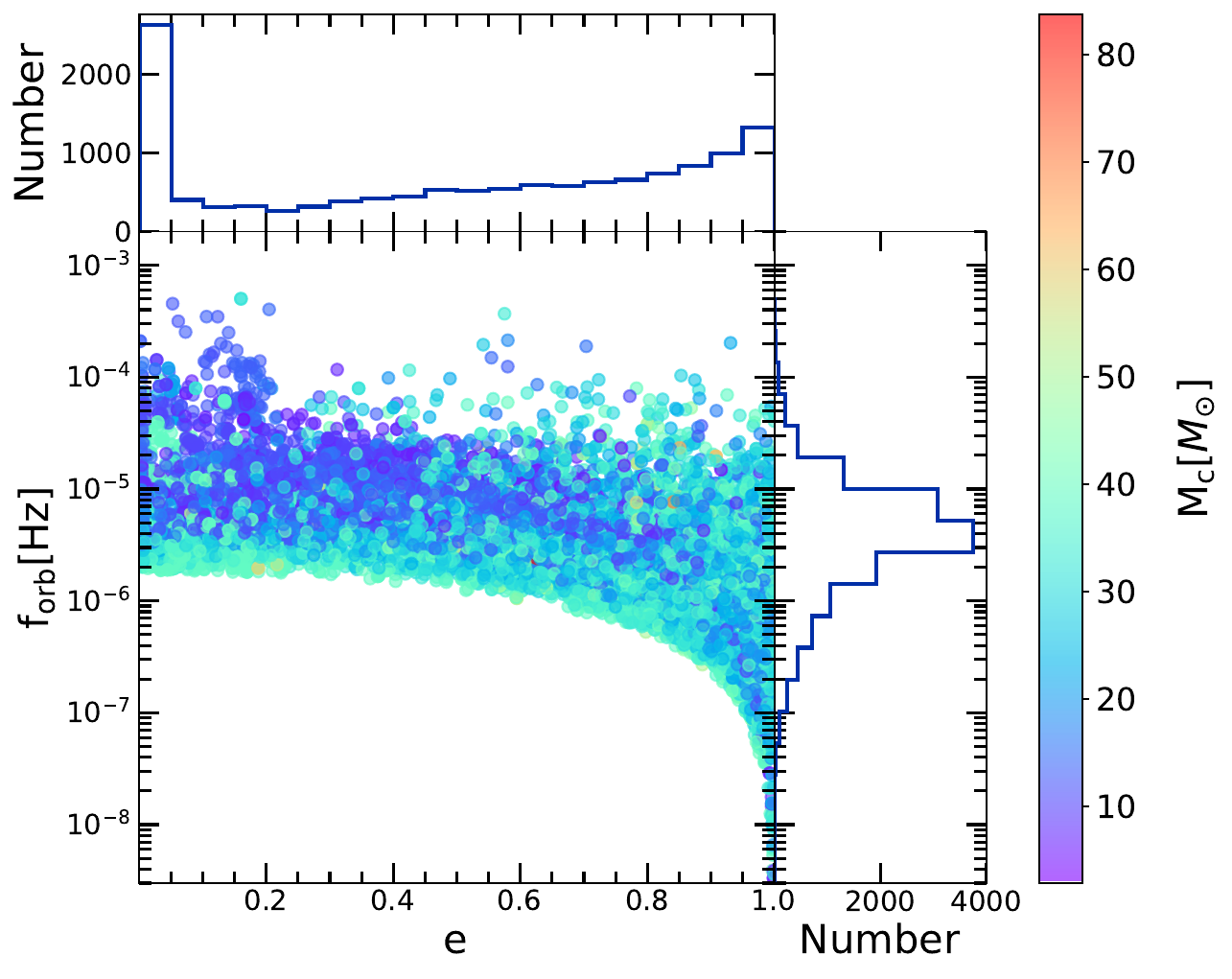}
\caption{
Orbital frequency ($\mathbf{f_{orb}}$) and eccentricity ($\mathbf{e}$) distribution of sBBHs escaping from GC models. Colors indicate the chirp masses of the sBBHs. The top and right-side panels show the distributions of eccentricity and orbital frequency, respectively.
}
\label{fig:f1}
\end{figure}

We select sBBHs that can merge within the Hubble time from 268 simulated cluster models in the MOCCA database, as described above. 
Figure~\ref{fig:f1} shows the distributions of eccentricity and orbital frequency for sBBHs at the moment they are ejected from their host GC. The colors indicate the chirp masses of the sBBH samples. Upon escape from the cluster, most sBBHs have orbital frequencies in the range $10^{-7}$\,Hz to $10^{-4}$\,Hz, with the peak centered around $5\times10^{-6}$\,Hz, as shown in the right-side panel of Figure~\ref{fig:f1}. 
Nearly $30\%$ of these sBBHs have high eccentricities ($e>0.8$) after ejection, and these highly eccentric sBBHs tend to have relatively lower orbital frequencies. A significant fraction (about $20\%$) of the escaping sBBHs are nearly circular.

Based on the original sample of escaping sBBHs extracted from the \textsc{MOCCA} models, we generate the entire cosmic sBBH sample. We assume that the cosmic GC formation rate follows~\citep{Mapelli2022}
\begin{equation}
\Psi_{\rm GC}(z) =  \mathcal{B}_{\rm GC}e^{-(z-z_{\rm GC})^2/(2\sigma_{\rm GC}^2)},
\label{eq:psigc}
\end{equation}
where $z_{\rm GC} = 3.2$ is the redshift of peak GC formation, $\sigma_{\rm GC} = 1.5$ is the standard deviation, and $\mathcal{B}_{\rm GC}$ is the normalization factor. We adopt $\mathcal{B}_{\rm GC} = 2\times10^{-4}\rm\, M_{\odot}\,Mpc^{-3}\,yr^{-1}$, consistent with the fiducial model in \citet{Mapelli2022}, and in agreement with \cite{ElBadry2019} and \cite{Reina2019}. Combining the GC formation rate and the mass function of GCs, we calculate the merger rate density of sBBHs escaping from GCs as
\begin{equation}
\mathcal{R}_{\rm GC} = \iint \frac{\Psi_{\rm GC}(z(\tau))}{\bar{M}}\,\Phi_{\rm GC}(M_{\rm GC})\, R(M_{\rm GC}, \tau-t_{\rm d})\, dM_{\rm GC}\, d\tau,
\label{eq:mrd}
\end{equation}
where $\bar{M}$ is the average mass for globular clusters which is set as $10^{5.6}\,\rm M_{\odot}$\citep{2010ARA&A..48..431P}, $\Phi_{\rm GC}$ is the mass function of GCs, following an evolved Schechter function~\citep{Jordan2007}:
\begin{equation}
\label{eq:phigc}
\Phi_{\rm GC} \propto (M_{\rm GC}+\Delta M)^{-2}\exp\left(-\frac{M_{\rm GC}+\Delta M}{M_{\rm cut}}\right),
\end{equation}
with $M_{\rm cut}$ as the high-mass cutoff for GCs and $\Delta M$ the average mass lost from each GC. At GC formation, we set $\Delta M = 0$, so Equation~\eqref{eq:phigc} becomes a standard Schechter function. For $M_{\rm cut}$, we use the maximum GC mass from the MOCCA models. $R(M_{\rm GC},\tau-t_{\rm d})$ is the merger rate density of escaping sBBHs from a GC with total mass $M_{\rm GC}$ and formation time $\tau$. The time delay $t_{\rm d}$ for each sBBH is
\begin{equation}
\label{eq:td}
t_{\rm d} = t_{\rm esc} + t_{\rm merger},
\end{equation}
where $t_{\rm esc}$ is the time from GC formation to sBBH escape, and $t_{\rm merger}$ is the time from escape to merger, which we estimate following \citep{Peters1964}:
\begin{equation}
\label{eq:dadt}
\left<\frac{da}{dt}\right> = -\frac{64}{5}\frac{G^3m_1m_2(m_1+m_2)}{c^5a^3}\frac{1+\frac{73}{24}e^2+\frac{37}{96}e^4}{(1-e^2)^{7/2}},
\end{equation}
\begin{equation}
\label{eq:dedt}
\left<\frac{de}{dt}\right> = -\frac{304}{15}\frac{G^3m_1m_2(m_1+m_2)}{c^5a^4}\frac{e(1+\frac{121}{304}e^2)}{(1-e^2)^{5/2}},
\end{equation}
where $G$ is the gravitational constant, $c$ is the speed of light, $a$ and $e$ are the semi-major axis and eccentricity, and $m_1$ and $m_2$ are the component masses of the sBBH. 

\begin{figure}
\centering
\includegraphics[width=\columnwidth]{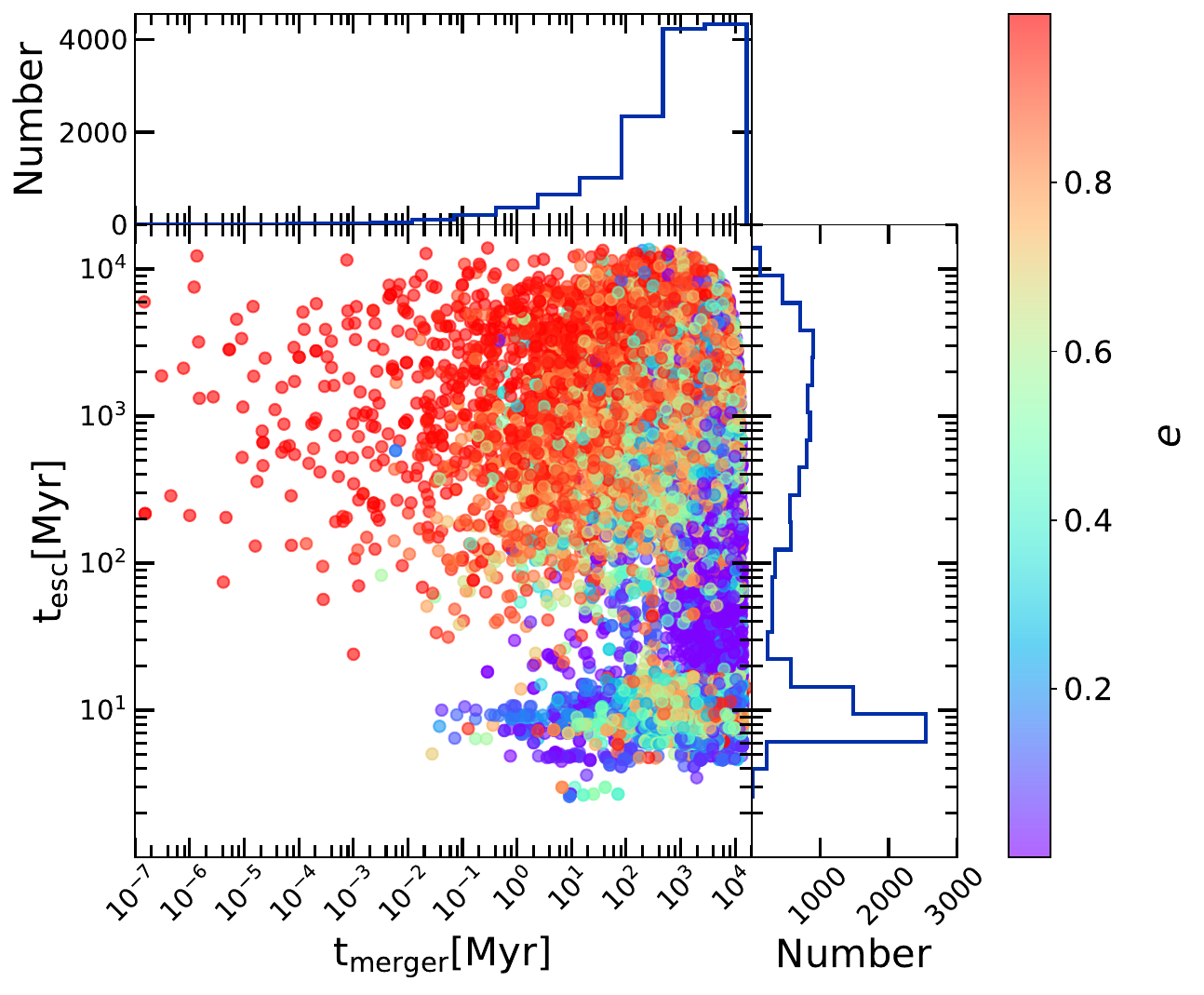}
\caption{
Distribution of escape timescale ($t_{\rm esc}$: the time from GC formation to sBBH ejection) and merger timescale ($t_{\rm merger}$: the time from escape to sBBH merger) for the sBBHs shown in Figure~\ref{fig:f1}. The top and right-side panels show the distributions of $t_{\rm merger}$ and $t_{\rm esc}$, respectively. The colors indicate the eccentricities of the sBBHs at escape.
}
\label{fig:f2}
\end{figure}

\begin{figure}
\includegraphics[width=\columnwidth]{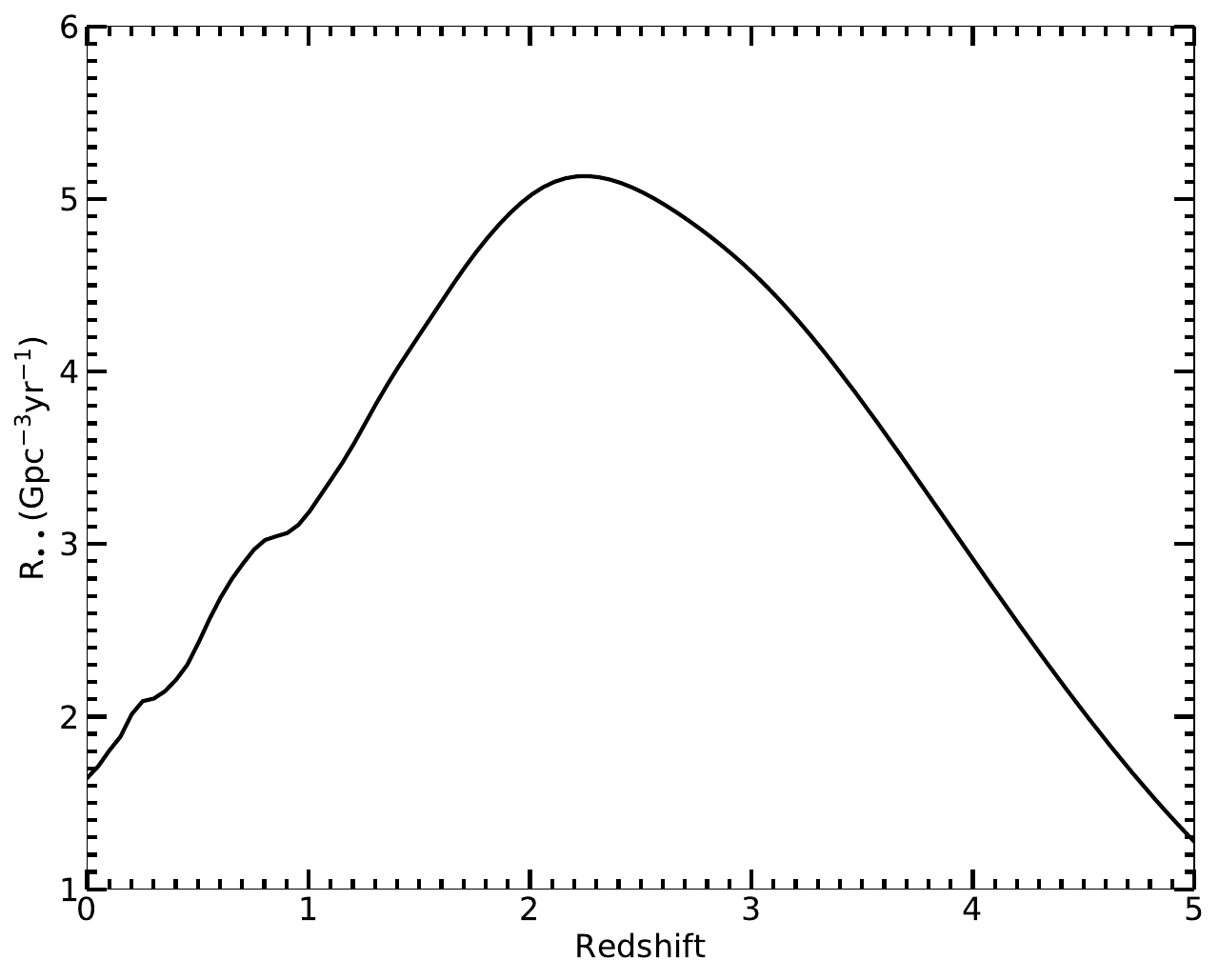}
\caption{
Merger rate density of escaping sBBHs originating from GCs.
}
\label{fig:f3}
\end{figure}

Figure~\ref{fig:f2} shows the escape timescale ($t_{\rm esc}$) and merger timescale ($t_{\rm merger}$) distributions for escaping sBBHs from the MOCCA models. The different colors in Figure~\ref{fig:f2} indicate the eccentricities of the sBBHs at the time of escape. Most escaping sBBHs have merger timescales between $10^3\,\rm Myr$ and $10^4\,\rm Myr$, and the escape timescale distribution peaks around $10\,\rm Myr$. sBBHs that escape quickly from the cluster tend to have lower eccentricities compared to those with longer escape timescales. This is reasonable, as sBBHs remaining longer in the cluster are subject to stronger dynamical interactions, resulting in higher eccentricities. Figure~\ref{fig:f3} presents the estimated merger rate density for escaping sBBHs originating from GCs. We find a local merger rate density of approximately $1.7$ Gpc$^{-3}$ yr$^{-1}$. For comparison, the total local merger rate density for sBBHs reported by the LVK collaboration after the first four observing runs is $\Rmrgbh(0) = 19^{+7}_{-5}$\citep{gwtc4}, indicating that escaping sBBHs from GCs contribute only a small fraction to the overall population. The merger rate density peaks at redshift $z \approx 2.2$, reaching around $5.13$ Gpc$^{-3}$ yr$^{-1}$. 
It is important to note that if we adopt the approach of \cite{2017Askar} the local merger rate density would increase to about $\bm12$ Gpc$^{-3}$ yr$^{-1}$. The reason for this difference comes from both a different assumption of the redshift evolution and mass distribution of the globular clusters.
Our estimate in this work is consistent with the range of local merger rates for sBBHs from GCs reported by \citet{Hong2018}.


\begin{figure*}
\centering
\includegraphics[width=\textwidth]{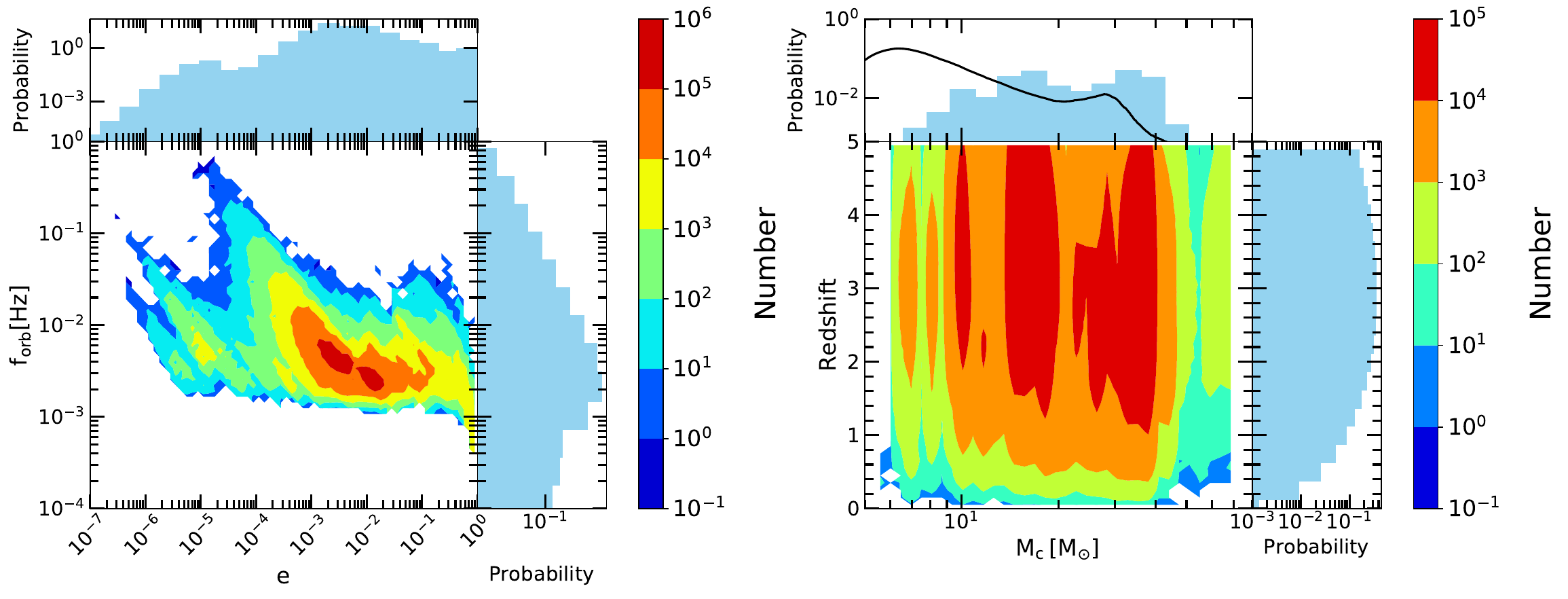}
\caption{
Parameter distributions of mock escaping sBBHs merging within $500$ years in the millihertz and decihertz bands. The left panel shows the initial eccentricities and orbital frequencies at the time of ejection from their host GC. The right panel shows intrinsic chirp mass ($M_{\rm c}$) and redshift for the sBBHs. Colors indicate the number of sources. The black solid line in the upper part of the right panel shows the observed chirp mass distribution of sBBHs from LVK constraints.
}
\label{fig:f4}
\end{figure*}

We generate the cosmic population of such sBBHs via the Monte Carlo method based on the merger rate density evolution described by Eq.(\ref{eq:mrd}), which is a wildly used method in this field \citep{ChenYu2020,ZhaoLu2023}, and the original samples of escaping sBBHs from the MOCCA models (see Fig.~\ref{fig:f1}). 
As our focus is on the millihertz and decihertz GW frequency bands, and considering computational cost and the sensitive frequency range for these sBBHs, we simulate only sources that can merge within $500$ years. Figure~\ref{fig:f4} shows the parameter distributions for the population of escaping sBBHs, including intrinsic chirp mass ($M_{\rm c} = m_1^{3/5}m_2^{3/5}/(m_1+m_2)^{1/5}$), redshift, eccentricity, and orbital frequency when their GW signals are firstly recorded by the millihertz space-based GW detectors such as LISA and Taiji during their observation period. Notice that we assume LISA and Taiji start the observation at the same time.
The left panel of Figure~\ref{fig:f4} shows that the orbital frequencies of escaping sBBHs range from $10^{-4}$\,Hz to about $0.8$\,Hz, with a peak near $2\times10^{-3}$\,Hz, which is where low-frequency GW detectors are most sensitive. Compared to the initial eccentricity distribution of sBBHs just after escaping from the MOCCA models, the distribution for those reaching the GW detection band is shifted, with a peak between $0.001$ and $0.01$.

After escape from the GC, the orbital separation of sBBHs shrinks and their eccentricities decay due to GW emission. Thus, when these sources enter the sensitivity window of low-frequency GW detectors, their eccentricities are much lower than at the time of escape. Some sBBHs escape with high orbital frequencies and high eccentricities, and thus can still have substantial eccentricity at GW observational frequencies. The right panel of Figure~\ref{fig:f4} shows the intrinsic chirp mass and redshift distributions for these mock sBBHs. The redshift distribution peaks at $z \sim 2$–$3$, reflecting the peak of GC formation and the relatively short time interval ($\sim 10^9$\,yr) between sBBH formation and merger. The intrinsic chirp mass distribution for escaping sBBHs is quite flat between $10\,M_{\odot}$ and $40\,M_{\odot}$, in contrast to the distribution for sBBHs formed via isolated evolution of massive binary stars\citep{ZhaoLu2021}, and similar to the location of the secondary peak in the BBH chirp mass distribution inferred from the GWTC-3 catalog \citep{GWTC3population} (black solid line in Fig.~\ref{fig:f4}).

This suggests that the high-mass peak in the observed sBBH chirp mass distribution could be due to sources originating from the dynamical channel. There are escaping sBBHs with chirp masses exceeding $60\, M_{\odot}$ in the MOCCA GC models, formed as second-generation sources after earlier mergers. This indicates that the mass function for sBBHs from GCs extends to higher masses than that for sBBHs formed through isolated evolution of massive binary stars \citep{ZhaoLu2021}, whose chirp mass distribution peaks at about $6 M_{\odot}$ and falls off rapidly above $40\,M_{\odot}$.

In addition to the mock sample described above, we generate nine more realizations following the same procedure to account for fluctuations due to the Monte Carlo method, particularly when the expected number of detections is small.

\section{Detectability of Eccentric sBBHs by Different GW Detectors} \label{sec:snr}

To investigate the detectability of eccentric, escaping sBBHs originating from GCs with different GW detectors, we calculate the signal-to-noise ratio (S/N) for each sBBH in the mock sample (see Section~\ref{sec:sample}). For eccentric sources, the GW emission frequency is not simply twice the orbital frequency; instead, it is a superposition of all integer multiples of the orbital frequency. The peak GW frequency for eccentric sources can be approximated as \citep{Randall2022}
\begin{equation}
\label{eq:fp}
f_{\rm p} \approx \frac{\sqrt{Gm}(1+e)^{\gamma}}{\pi [a(1-e^2)]^{3/2}},
\end{equation}
where $\gamma = 1.1954$. As eccentricity increases, the peak frequency deviates further from $2f_{\rm orb}$ (the value for circular binaries). The characteristic strain of the $n$th harmonic for an eccentric sBBH is calculated as
\begin{equation}
\label{eq:hcn}
h_{{\rm c},n}(f) = \frac{1}{\pi D}\sqrt{\frac{2G}{c^3}\frac{dE_n}{df_{\rm r}}},
\end{equation}
where $D$ is the comoving distance, $f_{\rm r} = n f_{\rm orb}$ is the GW frequency of the $n$th harmonic in the rest frame, and $dE_n/df_{\rm r}$ is the GW energy emitted at the $n$th harmonic in the source rest frame, given by \citep{1963Peters,2015Huerta}:
\begin{equation}
\frac{dE_n}{df_{\rm r}} = \frac{(2\pi G)^{2/3}M_{\rm c}^{5/3}}{3f_{\rm r}^{1/3}}\frac{g(n,e)}{n^{2/3}F(e)},
\label{eq:dEdf}
\end{equation}
where
\begin{eqnarray}
g(n,e)& = & \frac{n^4}{32} \left\{[J_{n-2}(ne)-2eJ_{n-1}(ne)+
\frac{2}{n}J_n(ne)\right. \nonumber \\
& &  + 2eJ_{n+1}(ne)-J_{n+2}(ne)]^2+(1-e^2) [J_{n-2}(ne)  \nonumber \\
& & \left. -2J_n(ne)+J_{n+2}(ne) ]^2+\frac{4}{3n^2}[J_n(ne)]^2 \right\},
\label{eq:gne}
\end{eqnarray}
and
\begin{equation}
F(e) = \frac{1+(73/24)e^2+(37/96)e^4}{(1-e^2)^{7/2}}.
\label{eq:fe}
\end{equation}
Here, $J_n$ are Bessel functions.

The signal-to-noise ratio (S/N) for an eccentric GW source can then be estimated by
\begin{equation}
\varrho^2 \approx 2\sum_{n=1}^{N}\int_{f_{\rm i}}^{f_{\rm f}} \frac{Qh_{c,n}^2(f)}{fS_n(f)}\frac{df}{f},
\label{eq:rho}
\end{equation}
where $S_n(f)$ is the non-sky-averaged power spectral density of a single detector, $f_{\rm i}$ is the initial frequency of the mock sBBH at the beginning of the observation, and $f_{\rm f}$ is given by
\begin{equation}
f_{\rm f} = \min[f_{\rm end}, f_{\rm ISCO}, f_{\rm detector}],
\end{equation}
where $f_{\rm end}$ is the GW frequency of the sBBH at the end of the observation, $f_{\rm ISCO} = 2.2M_{\odot}/[m_1(1+q)(1+z)]\, \rm{kHz}$ is the GW frequency at the innermost stable circular orbit (ISCO), and $f_{\rm detector}$ is the upper frequency cut-off of the detector. The parameter $Q$ in Equation~\eqref{eq:rho} accounts for the effects of source orientation and detector antenna pattern, which can be estimated as
\begin{equation}
\label{eq:Q}
Q = \sqrt{F_+^2\left(\frac{1+\cos^2\iota}{2}\right)^2+F_{\times}^2\cos^2\iota},
\end{equation}
where $F_+$ and $F_{\times}$ are the detector antenna pattern functions:
\begin{equation}
\label{eq:F_+}
F_+ = \frac{1}{2} \left(1+\cos^2 \theta \right) \cos 2\phi \cos 2\psi - \cos \theta \sin 2\phi \sin 2\psi,
\end{equation}
and
\begin{equation}
\label{eq:F_cross}
F_{\times} = \frac{1}{2}\left(1+\cos^2 \theta \right) \cos 2\phi \sin 2\psi + \cos \theta \sin 2\phi \cos 2\psi.
\end{equation}
Here, $\theta$, $\phi$, and $\psi$ are the polar angle, azimuthal angle, and polarization angle describing the sBBH's orientation in the detector frame, respectively, and $\iota$ is the inclination angle between the sBBH angular momentum and the line of sight from the detector. The S/N of an sBBH detected by multiple GW detectors can be estimated as
\begin{equation}
\label{eq:snrmulti}
\varrho^2 = \sum_{j=1}^{n}\varrho_j^2,
\end{equation}
where $j$ indexes the detectors and $n$ is the total number of detectors considered. To calculate the S/N for an sBBH merger event, we transform the parameters ($\theta$, $\phi$, $\psi$) in the detector frame to the corresponding angles ($\theta_{\rm S}$, $\phi_{\rm S}$, $\theta_{\rm L}$, $\phi_{\rm L}$) in the ecliptic coordinate frame following \citet{Cutler1998}.

\begin{table*}
\centering
\caption{
Expected number of detectable, escaping sBBHs from GCs in $4$-year observations by low-frequency GW detectors (LISA, Taiji, and LT), middle-frequency GW detectors (bAMIGO, AMIGO, and eAMIGO), and for multiband sBBHs detected by joint observations with both low- and middle-frequency detectors.
}
\label{tab:number}
\begin{tabular}{l|cccccccccc|ccc}  
\hline		
\hline
\multirow{2}{*}{GW detector}  & \multicolumn{10}{|c|}{realization} & \multirow{2}{*}{Median Value} & \multirow{2}{*}{Mean Value} & \multirow{2}{*}{STD} \\  
& $\rm r1$ & $\rm r2$ & $\rm r3$ & $\rm r4$ & $\rm r5$ & $\rm r6$ & $\rm r7$ & $\rm r8$ & $\rm r9$ & $\rm r10$  \\ \hline
{LISA} & $0$ & $0$ & $2$ & $2$ & $1$ & $1$ & $1$ & $0$ & $1$ & $0$ & $1.0$ & $0.8$ & $0.7$ \\ \hline
{Taiji} & $13$ & $9$ & $12$ & $14$ & $11$ & $15$ & $8$ & $11$ & $11$ & $12$ & $11.5$ & $11.6$ & $2.0$ \\ \hline
{bAMIGO} & $0$ &$0$ & $2$ & $1$ & $2$ & $1$ & $2$ & $0$ & $1$ & $0$ & $1.0$ & $0.9$ & $0.8$ \\ \hline
{AMIGO} & $8$ & $8$ & $11$ & $6$ & $7$ & $7$ & $8$ & $9$ & $8$ & $7$ & $8.0$ & $7.9$ & $1.3$  \\ \hline
{eAMIGO} & $11$ & $10$ & $12$ & $8$ & $8$ & $8$ & $10$ & $12$ & $9$ & $10$ & $10.0$ & $9.8$ & $1.5$ \\ \hline
{LT} & $16$ & $13$ & $14$ & $18$ & $13$ & $20$ & $12$ & $19$ & $13$ & $16$ & $15.0$ & $15.4$ & $2.7$ \\ \hline
{LISA-bAMIGO} & $1$ & $0$ & $3$ & $2$ & $1$ & $2$ & $3$ & $0$ & $2$ & $1$ & $1.5$ & $1.5$ & $1.0$ \\ \hline
{LISA-AMIGO} & $2$ & $0$ & $3$ & $2$ & $3$ & $2$ & $4$ & $0$ & $2$ & $1$ & $2.0$ & $1.9$ & $1.2$  \\ \hline
{LISA-eAMIGO} & $2$ & $0$ & $3$ & $2$ & $3$ & $2$ & $4$ & $0$ & $2$ & $1$ & $2.0$ & $1.9$ & $1.2$ \\ \hline
{Taiji-bAMIGO} & $2$ & $2$ & $6$ & $3$ & $4$ & $3$ & $3$ & $3$ & $4$ & $3$ & $3.0$ & $3.3$ & $1.1$ \\ \hline
{Taiji-AMIGO} & $22$ & $17$ & $20$ & $20$ & $16$ & $20$ & $16$ & $23$ & $19$ & $26$ & $20.0$ & $19.9$ & $3.0$  \\ \hline
{Taiji-eAMIGO} & $26$ & $19$ & $25$ & $27$ & $22$ & $21$ & $17$ & $28$ & $22$ & $29$ & $23.5$ & $23.6$ & $3.8$ \\ \hline
{LT-bAMIGO} & $2$ & $3$ & $6$ & $3$ & $4$ & $3$ & $3$ & $3$ & $4$ & $3$ & $3.0$ & $3.4$ & $1.0$\\ \hline
{LT-AMIGO} & $22$ & $20$ & $20$ & $20$ & $16$ & $21$ & $17$ & $24$ & $19$ & $27$ & $20.0$ & $20.6$ & $3.0$ \\ \hline
{LT-eAMIGO} & $26$ & $22$ & $25$ & $29$ & $23$ & $23$ & $18$ & $29$ & $22$ & $31$ & $24.0$ & $24.8$ & $3.8$  \\ \hline
\end{tabular}
\begin{flushleft}
\footnotesize{Note: The first column lists the GW detector name or joint combinations. ‘LT’ represents the LISA-Taiji network. Columns two to eleven show results from $10$ realizations of the number of detectable sBBHs ejected from GCs. The last three columns give the median, mean, and standard deviation across the $10$ realizations.}
\end{flushleft}
\end{table*}

For the mock samples of escaping sBBHs described in Section~\ref{sec:sample}, we randomly assign their orientation parameters, assuming a uniform distribution on the sky. Specifically, $\phi_{\rm S}$ and $\phi_{\rm L}$ are uniformly distributed in $[0,2\pi)$, while $\cos\theta_{\rm S}$ and $\cos\theta_{\rm L}$ are uniformly distributed in $[-1,1]$. We then estimate the S/Ns for all mock sBBHs in the $10$ realizations generated, for low-frequency detectors, middle-frequency detectors, and joint observations of low- and middle-frequency detectors. For the low-frequency band, we consider LISA, Taiji, and the LISA-Taiji network (hereafter LT). The LISA and Taiji sensitivity curves are adopted from \citet{Robson2019} and \citet{Wanggang2020}, respectively. For the middle-frequency band, we consider the AMIGO project with three different sensitivity curves as described in \citet{Ni2022}.

Table~\ref{tab:number} lists the expected number of detectable, escaping sBBHs originating from GCs over a $4$-year observation period for LISA, Taiji, LT, bAMIGO, AMIGO, eAMIGO, and for various combinations of low- and middle-frequency detectors. We assume that an sBBH is detected in single-band observations if $\varrho \geq 8$. For multiband observations, the detection criterion is that the S/N in both the low- and middle-frequency band detectors exceeds $5$. 

LISA/Taiji may detect around $0.8$/$11.6$ escaping sBBHs after a $4$-year mission. LT slightly increases this number to about $15.4$, although the improvement is not significant since LISA is less sensitive to these sources. The detection numbers for AMIGO, with its three different sensitivity levels, vary considerably: the least sensitive version could detect about $0.9$ sBBHs, while the most sensitive could detect up to $9.8$ sBBHs. For multiband observations, the joint detection by LISA-bAMIGO may yield about $1.5$ sBBHs, and the LT-eAMIGO combination—which is the most sensitive configuration considered here—could detect around $24.8$ sBBHs. The number of multiband detections by low- and middle-frequency detectors is not simply the sum of those for individual detectors, due to the different detection criteria.

\begin{figure*}
\centering
\includegraphics[width=\textwidth]{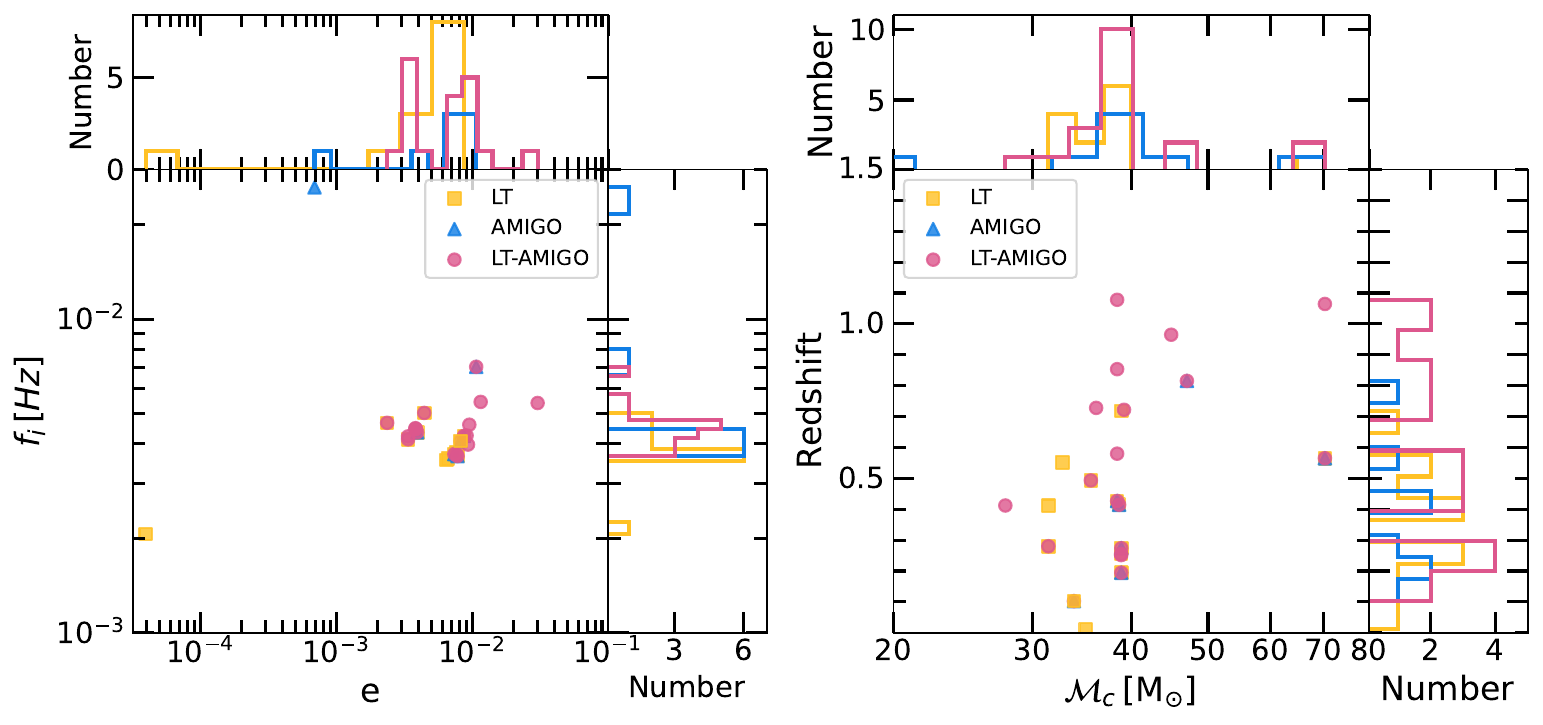}
\caption{
Parameter distributions of detectable escaping sBBHs in realization $9$ for LT, AMIGO, and their joint observations. The left panel shows the initial eccentricities and initial orbital frequencies of these sBBHs at the start of observations for different GW detectors, along with their corresponding distributions. The right panel shows the two-dimensional distribution of escaping sBBHs in the plane of redshifted chirp mass ($\mathcal{M}_{\rm c}=(1+z)M_{\rm c}$) and redshift, as well as one-dimensional histograms for these two parameters. Yellow squares represent sBBHs detectable by LT ($\varrho_{\rm LT} \geq 8$), blue triangles those detectable by AMIGO ($\varrho_{\rm AMI} \geq 8$), and red circles the multiband sBBHs detectable by LT-AMIGO ($\varrho_{\rm LT} \geq 5$ and $\varrho_{\rm AMI} \geq 5$).
}
\label{fig:f5}
\end{figure*}

Figure~\ref{fig:f5} shows the parameter distributions of escaping sBBHs detectable by the low-frequency GW detector (the LT network), the middle-frequency GW detector (AMIGO), and by joint observations with both (LT-AMIGO) during $4$ years of observation in realization $9$. This particular realization was chosen because it is representative of the median number of detectable sources for the various GW detectors and their combinations, as shown in Table~\ref{tab:number}. The left panel of Figure~\ref{fig:f5} presents the initial eccentricities and orbital frequencies of the detectable sBBHs observed by the LT network (yellow squares), AMIGO (blue triangles), and LT-AMIGO (red circles).

The peaks of the initial eccentricity distributions for sources detected by LT and LT-AMIGO are centered around $7\times10^{-3}$, similar to the peak (around $3\times10^{-3}$) of the eccentricity distribution of the entire BBH sample in Figure~\ref{fig:f4}. The initial eccentricity distribution for sBBHs detected by AMIGO is also centered around $7\times10^{-3}$. Although some escaping sBBHs have high eccentricities at the time of escape, their eccentricities decrease significantly by the time they reach the low- and middle-frequency bands. LT-AMIGO detects one sBBH with an initial eccentricity exceeding $0.02$, but the initial eccentricities of most sources detected by LT, AMIGO, and LT-AMIGO are less than $0.01$. The right panel of Figure~\ref{fig:f5} shows the distributions of redshifted chirp mass and redshift for these detectable sBBHs.

Most sBBHs detected by LT, AMIGO, or LT-AMIGO have redshifted chirp masses in the range $20$–$50\,M_{\odot}$, except for two sBBHs detected by LT-AMIGO, which have redshifted chirp masses of $70\,M_{\odot}$. The redshift distribution of sBBHs detected by LT-AMIGO extends to higher redshifts than those detected by AMIGO or LT alone. Most sBBHs detected by either LT or AMIGO individually are nearby sources, with redshifts less than $0.6$. LT-AMIGO, however, can detect sBBHs over a redshift range of $0$–$1.1$.

\begin{figure*}
\centering
\includegraphics[width=\textwidth]{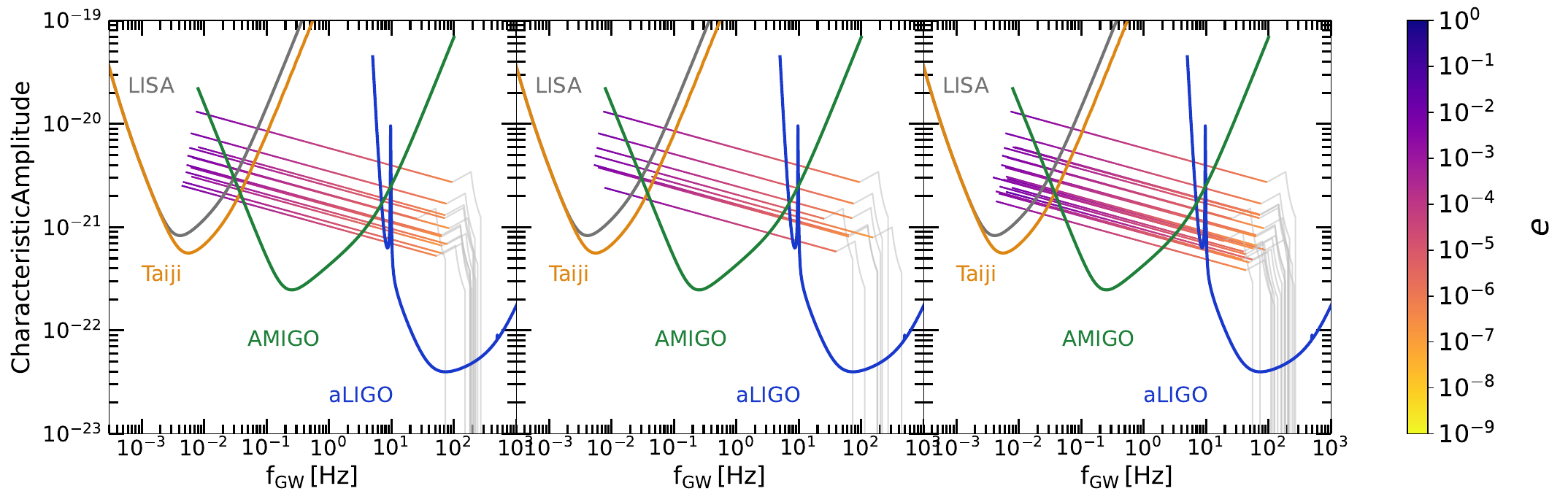}
\caption{
Characteristic amplitude tracks of the mock escaping sBBHs in realization $9$ detectable by LT alone (left panel, $\varrho_{\rm LT} \geq 8$), AMIGO alone (middle panel, $\varrho_{\rm AMI} \geq 8$), and LT-AMIGO (right panel, $\varrho_{\rm LT} \geq 5$ and $\varrho_{\rm AMI} \geq 5$) during a continuous $4$-year observation. Color indicates the evolution of eccentricity during the inspiral phase, while light gray tracks correspond to the merger and ringdown stages. The gray, orange, green, and blue curves represent the sensitivity curves of LISA, Taiji, AMIGO, and the aLIGO design, respectively.
}
\label{fig:f6}
\end{figure*}

Figure~\ref{fig:f6} shows the evolution of the GW characteristic strain amplitude for the peak harmonic ($\max(h_{{\rm c},n})$) for the eccentric, escaping mock sBBHs in realization $9$ that are detectable by the low-frequency detector(s), middle-frequency detector, and by joint observations. The sBBHs with $\varrho_{\rm LT} \geq 8$ are shown in the left panel of Figure~\ref{fig:f6}. Since all these detectable sources have eccentricities below $0.1$, their characteristic strain amplitude evolution tracks are similar to those of circular binaries.

During the inspiral stage, the relationship between eccentricity and orbital frequency follows \citep{2007Enoki}:
\begin{equation}
\label{eq:f-e}
\frac{f_{\rm orb}}{f_{0}} = \left[\frac{1-e_0^2}{1-e^2}\left(\frac{e}{e_0}\right)^{12/19}\left(\frac{1+\frac{121}{304} e^2}{1+\frac{121}{304}e_0^2}\right)^{870/2299}\right]^{-3/2}.
\end{equation}
Since the residual eccentricity is nearly zero at the end of the inspiral, we estimate the characteristic strain amplitude for the merger and ringdown stages using the analytic fits from \citet{2008Ajith} (see light gray tracks in Figure~\ref{fig:f6}). All $13$ sBBHs detected by LT will merge within $4$ years.

The middle panel of Figure~\ref{fig:f6} shows the evolution for sBBHs detected by AMIGO ($\varrho_{\rm AMI} \geq 8$). The right panel illustrates the increased number of multiband sBBHs detected by LT-AMIGO, all of which are also close to circular orbits.

\section{Parameter Estimation Precision for the Eccentric sBBHs} \label{sec:PE}

Eccentricity is an important parameter for distinguishing the formation channels of sBBHs. In this section, we analyze the parameter estimation accuracy for escaping sBBHs detected by GW observatories in different frequency bands. Although the residual eccentricity in the millihertz and decihertz bands for the escaping sBBHs studied above is quite small, it can still provide useful information in parameter estimation analyses. To estimate the measurement uncertainties of sBBH parameters, we adopt the traditional Fisher Information Matrix (FIM) method, which is widely used in previous studies \citep{Cutler1994, Grimm2020, LiuShao2020, ChenLu2021, ZhaoLu2023}. Under the assumptions of Gaussian noise and the high S/N limit, the covariance matrix of the GW source parameters is the inverse of the Fisher matrix, providing a quick estimate of the precision of GW source parameters. While these assumptions may not always hold in practice, further discussion can be found in Section~\ref{sec:dis_con}. 

The Fisher matrix is defined as
\begin{equation}
\label{eq:Gamma}
\Gamma_{ab} =  \left(\left. \frac{\partial h}{\partial \Xi^{a}} \right| \frac{\partial h}{\partial \Xi^{b}}\right),
\end{equation}
where $h$ is the frequency-domain waveform of the GW source, and $\Xi$ denotes the set of source parameters. Since we focus on sBBHs originating from GCs, we take ten parameters into account: $d_{\rm L}$, $m_1$, $q$, $e_0$, $t_{\rm c}$, $\phi_{\rm c}$, $\theta_{\rm S}$, $\phi_{\rm S}$, $\theta_{\rm L}$, and $\phi_{\rm L}$. To investigate the effect of multiband observations (including low-, middle-, and high-frequency bands) on parameter estimation precision, we adopt the complete inspiral-merger-ringdown waveform model EccentricFD, which incorporates eccentricity, as implemented in {\bf PyCBC} \citep{2019PASP..131b4503B}.

For multiband GW observations or networks of multiple detectors operating in a single frequency band, the Fisher matrix can be calculated as
\begin{eqnarray}
\label{eq:Gamma2}
\Gamma_{ab} & = & \left(\left.\frac{\partial h}{\partial \boldsymbol{\Xi}^{a}} \right| \frac{\partial h}{\partial \boldsymbol{\Xi}^{b}}\right) \nonumber \\
& = & \sum_{j=1}^{n} 2\int_{f_{\rm i}}^{f_{\rm f}} \frac{\frac{\partial \tilde{h}_j^{*}(f)}{\partial \boldsymbol{\Xi}^{a}}\frac{\partial \tilde{h}_j(f)}{\partial \boldsymbol{\Xi}^{b}}+\frac{\partial \tilde{h}_j^{*}(f)}{\partial \boldsymbol{\Xi}^{b}}\frac{\partial \tilde{h}_j(f)}{\partial \boldsymbol{\Xi}^{a}}}{S_{{\rm n},j}(f)} df,
\end{eqnarray}
where $j$ indexes the independent detectors and $n$ is the total number of Michelson interferometers in the detector network. More details and technical settings can be found in \citet{ZhaoLu2023}.

The covariance matrix is then given by
\begin{equation}
\Sigma = \left\langle\delta \Xi^{a} \delta \Xi^{b}\right\rangle = \left(\Gamma^{-1}\right)^{ab},
\end{equation}
and the uncertainty in each sBBH parameter estimate is
\begin{equation}
\Delta \Xi^{a}=\sqrt{\left(\Gamma^{-1}\right)^{aa}}.
\end{equation}
For the angular resolution $\Delta\Omega$, we specifically consider the $90\%$ confidence level, which can be obtained by \citep{Cutler1994,Barack2004,WenChen2010}
\begin{align}
\Delta\Omega_{{\rm X}\%} =\; & -2\pi|\sin\theta_{\rm S}| \notag \\
& \times \sqrt{(\Delta\theta_{\rm S}\Delta\phi_{\rm S})^2 
    - \left\langle\Delta\theta_{\rm S}\Delta\phi_{\rm S}\right\rangle^2}
    \ln(1-{\rm X}/100)
\end{align}
where ${\rm X}\%$ represents the confidence level (we adopt ${\rm X} = 90$ throughout this paper). $\Delta\theta_{\rm S}$, $\Delta\phi_{\rm S}$, and $\left\langle\Delta\theta_{\rm S}\Delta\phi_{\rm S}\right\rangle$ are obtained from the covariance matrix. 

\begin{figure*}
\centering
\includegraphics[width=\textwidth]{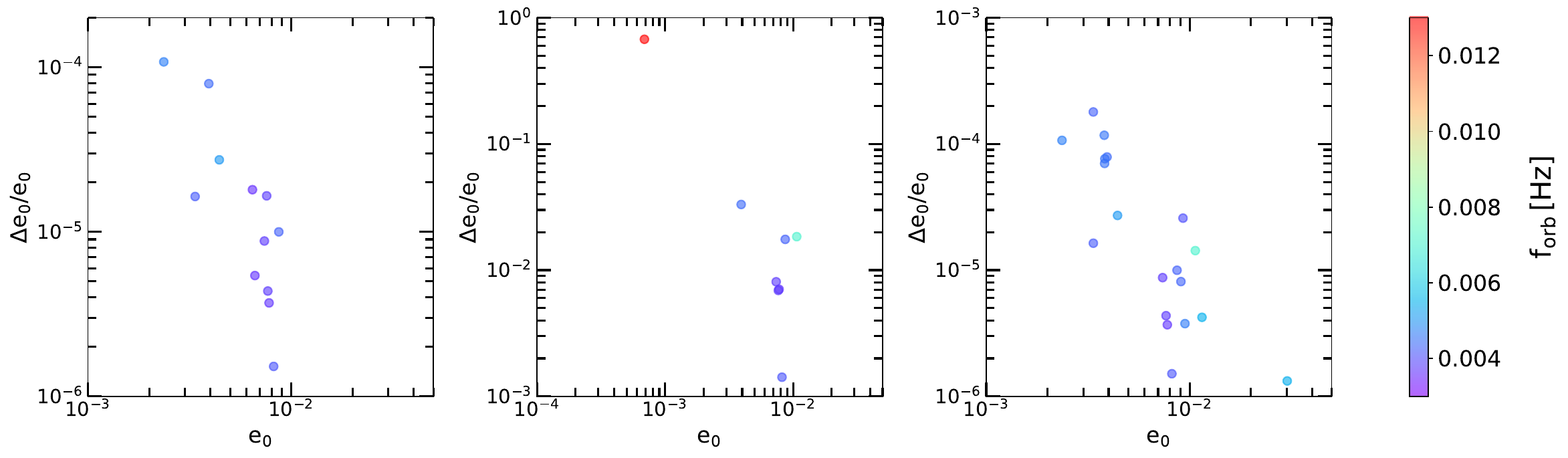}
\caption{Measurement precision of initial eccentricity{\textbf{($\mathbf{e_0}$), which means the eccentricity value at the beginning of GW observation,}} for escaping sBBHs detected by the LT network (left panel), AMIGO (middle panel), and joint LT-AMIGO observations (right panel). Colors represent the initial orbital frequencies for these binaries at the beginning of the observations.}
\label{fig:f7}
\end{figure*}

\begin{figure*}
\centering
\includegraphics[width=\textwidth]{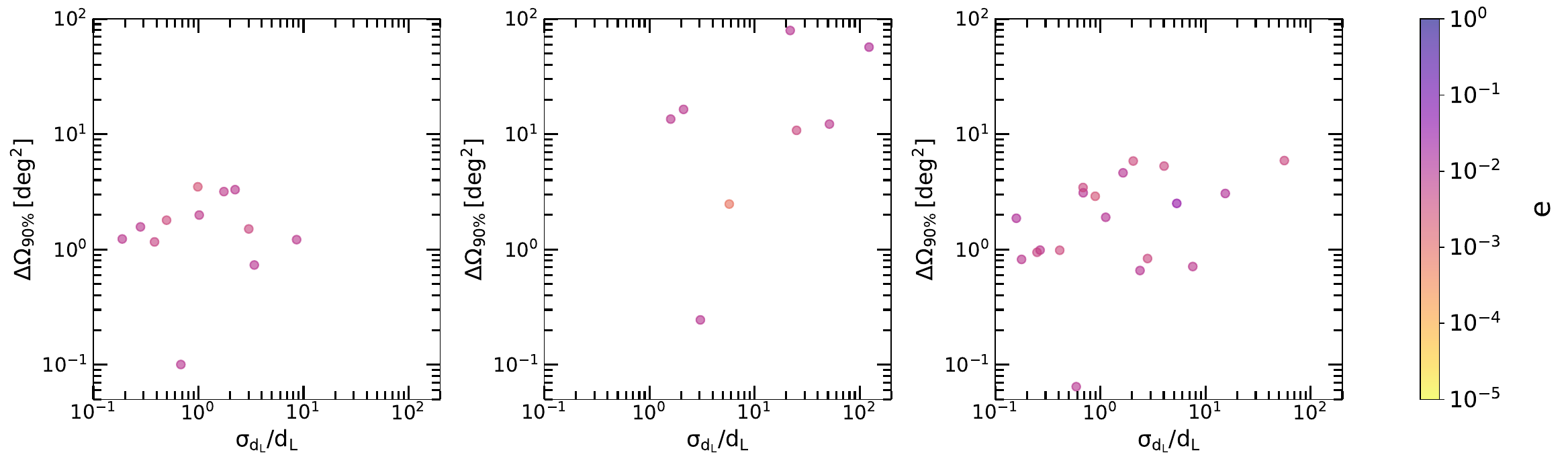}
\caption{Measurement precision for relative luminosity distance and sky localization of escaping sBBHs detected by the LT network (left panel), AMIGO (middle panel), and joint LT-AMIGO observations (right panel). Colors represent the initial eccentricities of these binaries at the beginning of the observations.}
\label{fig:f8}
\end{figure*}

Figure~\ref{fig:f7} shows the relative measurement precision of initial eccentricity for detectable escaping sBBHs as measured by LT, AMIGO, and LT-AMIGO. We find that LT can constrain the initial eccentricity with relative precision below $10^{-4}$ for most detectable sources (see the left panel of Fig.~\ref{fig:f7}). Sources with larger initial eccentricities tend to be constrained with higher precision; for example, most sBBHs with initial eccentricity greater than $5\times10^{-3}$ can be measured with $\Delta e_0/e_0 < 10^{-5}$. For sBBHs observed by AMIGO (middle panel), the relative eccentricity errors range from $10^{-3}$ to $0.7$, which are larger than the values obtained from LT. This is because eccentricity decreases rapidly as the sources evolve to higher-frequency bands. For those multiband sBBHs that can be observed by both LT and AMIGO, the relative measurement precision of the initial eccentricity is $10^{-6}$--$2\times10^{-4}$ (right panel of Fig.~\ref{fig:f7}). Although the combination of LT and AMIGO improves the measurement precision, the dominant contribution to initial eccentricity estimation comes from the low-frequency GW observations by LT.

Figure~\ref{fig:f8} shows the sky localization errors and relative luminosity distance errors for escaping sBBHs observed by LT, AMIGO, and LT-AMIGO, respectively. The low-frequency detector LT can localize sBBHs to sky areas of $0.1$--$4$\,deg$^2$ (left panel of Fig.~\ref{fig:f8}), with relative errors in luminosity distance ranging from $0.1$ to $10$. The middle panel of Figure~\ref{fig:f8} shows that AMIGO can localize sBBHs to sky areas of $0.2$--$100$\,deg$^2$, and measure luminosity distance with relative errors of $1$--$200$. Compared to LT, the localization and luminosity distance measurement precision achieved by AMIGO is less accurate. For multiband sBBHs, LT-AMIGO may localize them to sky areas of $0.05$--$5$\,deg$^2$ and measure luminosity distance with relative errors of $\sim 0.1$--$60$. The combination of low- and middle-frequency band observations therefore improves both the localization area and luminosity distance measurements compared to single-band observations.

\begin{figure*}
\centering
\includegraphics[width=\textwidth]{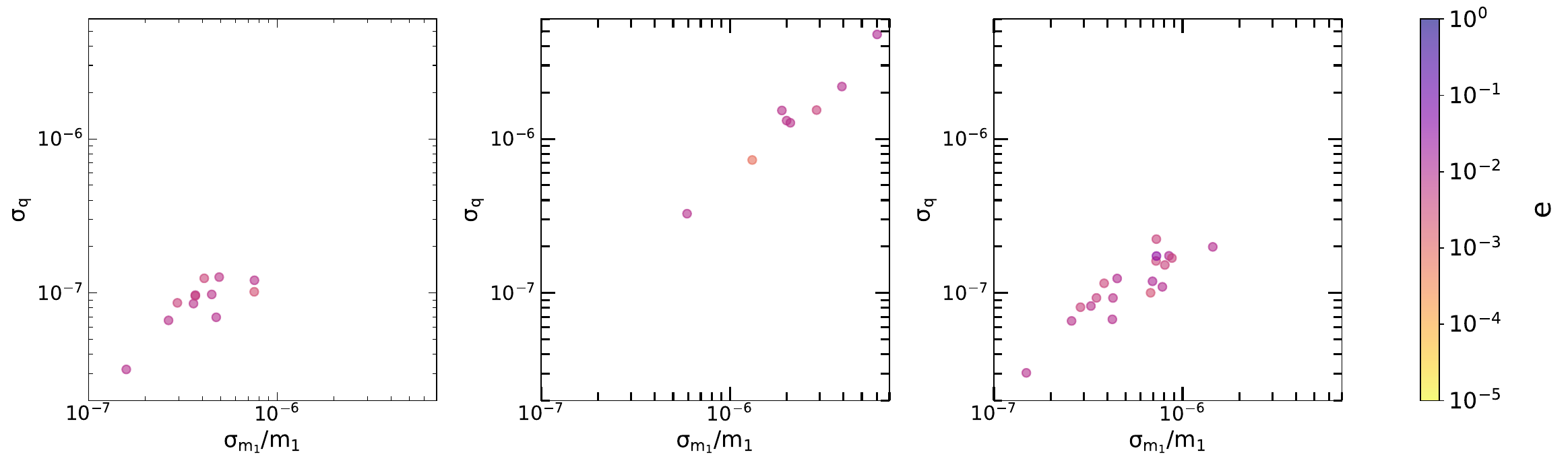}
\caption{
Relative errors of the primary mass and mass ratio for escaping sBBHs detected by the LT network (left panel), AMIGO (middle panel), and joint LT-AMIGO observations (right panel). Colors represent the initial eccentricities of these binaries at the beginning of the observations.
}
\label{fig:f9}
\end{figure*}

Figure~\ref{fig:f9} shows the measurement precision for primary mass and mass ratio for escaping sBBHs detectable by LT, AMIGO, and LT-AMIGO, respectively. LT can measure the primary mass with relative errors of $10^{-7}$--$8\times10^{-7}$, and measure the mass ratio with errors of $3\times10^{-8}$--$2\times10^{-7}$. The middle panel of Figure~\ref{fig:f9} shows that AMIGO can measure the primary mass with relative errors of $6\times10^{-7}$--$6\times10^{-6}$ and the mass ratio with errors of $3\times10^{-7}$--$5\times10^{-6}$. The estimation results for $m_1$ and $q$ by both AMIGO and LT are highly accurate, with LT achieving slightly better constraints since the sBBHs detected by LT have lower frequencies and experience more cycles during their inspiral. For multiband sBBHs, LT-AMIGO can measure the primary mass with relative errors of $10^{-7}$--$2\times10^{-6}$ and the mass ratio with errors of $3\times10^{-8}$--$3\times10^{-7}$. It is clear that joint observations with low- and middle-frequency detectors further improve the precision of measurements for the primary mass and mass ratio of sBBHs, as shown in the right panel of Figure~\ref{fig:f9}.

Finally, we summarize the measurement precision results for initial eccentricity, sky localization, luminosity distance, mass ratio, and primary mass of escaping sBBHs observed by low-frequency, middle-frequency, and joint low- and middle-frequency GW detectors in Table~\ref{tab:precision}.

\section{Conclusions and Discussion}
\label{sec:dis_con}

In this study, we investigated eccentric, escaping sBBHs from GCs and assessed their detectability and parameter estimation precision using low- and middle-frequency GW observations, as well as multiband observations that combine these bands. We generated the cosmic population of escaping sBBHs originating from GC models simulated with the MOCCA code, and used Monte Carlo methods to estimate the numbers of detectable sBBHs with low-frequency detectors (LISA, Taiji, LT), middle-frequency detectors (bAMIGO, AMIGO, eAMIGO), and their combinations. We also analyzed the measurement precision for the eccentricity, localization, and other parameters of these sBBHs.

\begin{table*}
\centering
\caption{Measurement precision for the initial eccentricity, sky localization, luminosity distance, mass ratio, and primary mass of escaping sBBHs detected by LT, AMIGO, and LT-AMIGO, respectively.}
\label{tab:precision}
\begin{tabular}{lccccc}
\hline
\hline
     & $\Delta e_0/e_0$ & $\Delta \Omega_{90\%}(\rm deg^2)$ & $\sigma_{\rm d_{L}}/d_{L}$ & $\sigma_{\rm m_1}/{\rm m_1}$ & $\sigma_{\rm q}$\\ \hline
{LT} & $10^{-6}-2\times10^{-4}$ & $0.1-3$ & $0.1-10$ & $10^{-7}-8\times10^{-7}$ & $3\times10^{-8}-2\times10^{-7}$\\ \hline
{AMIGO} & $10^{-3}-0.7$ & $0.2-100$ & $1-200$ & $5\times10^{-7}-7\times10^{-6}$ & $3\times10^{-7}-5\times10^{-6}$\\ \hline
{LT-AMIGO} & $10^{-6}-2\times10^{-4}$ & $0.06-6$ & $0.1-6$ & $10^{-7}-2\times10^{-6}$ & $3\times10^{-8}-3\times10^{-7}$\\ \hline
\end{tabular}
\end{table*}

We find that LISA and Taiji could detect approximately $0.8\pm0.7$ and $11.6\pm2.0$ escaping sBBHs with S/N $\geq 8$, respectively, based on ten mock realizations and a continuous $4$-year observation period. The LT network could detect $15.4\pm2.7$ sBBHs, a significant improvement over LISA alone. For the three levels of middle-frequency detector sensitivity, bAMIGO, AMIGO, and eAMIGO could detect $0.9\pm0.8$, $7.9\pm1.3$, and $9.8\pm1.5$ sBBHs with S/N $\geq 8$, respectively. Combining the low- and middle-frequency detectors, LT-AMIGO could detect $20.6\pm3.0$ sBBHs with S/N $\geq 5$ in both bands within $4$ years. The joint observation with LT-AMIGO increases the number of detections compared to either LT or AMIGO alone, since LT and AMIGO have similar sensitivities and a different detection criterion is adopted for multiband observations.

Although a fraction of escaping sBBHs have high eccentricities (greater than $0.8$) at the time of escape, the sources detectable by LT and AMIGO exhibit small eccentricities (less than $0.04$). The initial frequency distributions of most detectable escaping sBBHs for LT, AMIGO, and LT-AMIGO are centered around $4\times10^{-3}\,\rm{Hz}$. We find that the redshifted chirp mass of most sBBHs detected by LT, AMIGO, and LT-AMIGO lies in the range $30$–$40\,\rm{M_{\odot}}$, and most detected sources are nearby, with redshift less than $1$. Compared to the detectable sBBHs originating from the isolated evolution of massive bianry stars, whose redshifted chirp mass distribution peaks between $20$–$30\,\rm M_{\odot}$\citep{ZhaoLu2023}, the detectable escaping sBBHs from GCs have somewhat higher chirp masses. The difference for the detectable chirp mass for the two channels results from the intrinsic mass distribution for two populations and the difference of redshift distributions for them. Although many sBBHs originating from GCs have high eccentricities at the time of escape, those detected by LT, AMIGO, or LT-AMIGO do not, since eccentricities decrease rapidly due to circularization from GW emission.

The relative measurement precision of initial eccentricities for the detectable escaping sBBHs is about $10^{-6}$–$2\times10^{-4}$ for LT and LT-AMIGO, and $10^{-3}$–$0.7$ for AMIGO. The precision with which eccentricity can be measured depends on its value: sources with higher eccentricities tend to have more accurately measured initial eccentricities. The low-frequency GW detector (LT) achieves better constraints on the initial eccentricity than the middle-frequency detector (AMIGO), since eccentricity decreases rapidly as the orbital frequency increases during the inspiral. The most sensitive frequency band of AMIGO is around the decihertz range ($0.1\,\rm Hz$), where most sBBHs have nearly circular orbits. Therefore, LT provides better estimates of the initial eccentricities. For multiband sBBHs, joint observation with LT-AMIGO significantly improves the precision of initial eccentricity estimates compared to AMIGO alone.

The measurement precision of sky localization for escaping sBBHs detected by LT, AMIGO, and LT-AMIGO is about $0.1$–$3\,\rm deg^2$, $0.2$–$100\,\rm deg^2$, and $0.06$–$6\,\rm deg^2$, respectively. The measurement precision for luminosity distance is about $0.1$–$10$ for LT, $1$–$200$ for AMIGO, and $0.1$–$6$ for LT-AMIGO. The relative error in primary mass measurement is about $10^{-7}$–$8\times10^{-7}$ for LT, $5\times10^{-7}$–$7\times10^{-6}$ for AMIGO, and $10^{-7}$–$2\times10^{-6}$ for LT-AMIGO. The measurement precision for the mass ratio is $3\times10^{-8}$–$2\times10^{-7}$ for LT, $3\times10^{-7}$–$5\times10^{-6}$ for AMIGO, and $3\times10^{-8}$–$3\times10^{-7}$ for LT-AMIGO. In general, combining low- and middle-frequency band observations improves the uncertainties for both intrinsic parameters and localization, compared to single-band observations.

\subsection{Caveats and Limitations}\label{sec:caveats}

While our study provides new insights into the multiband GW observations of eccentric sBBHs originating from GCs, there are several caveats and limitations that should be kept in mind when interpreting our results.

First, our analysis relies on a suite of star cluster models simulated with the MOCCA code, which is well suited for modeling the long-term dynamical evolution of large-\(N\), spherically symmetric star clusters. This approach is computationally efficient and has been validated against direct $N$-body simulations for a wide range of cluster models. However, MOCCA necessarily adopts several simplifying assumptions regarding the dynamical environment, stellar evolution, and binary interactions. For example, the underlying potential is assumed to be spherically symmetric, and the external tidal field is treated with a simplified prescription. Such approximations may not capture the full complexity of real clusters, particularly those exhibiting significant rotation, substructure, or non-spherical geometries.

The initial conditions adopted for our cluster models span a broad range in mass, density, binary fraction, and metallicity, and the present-day properties of our simulated clusters are comparable to those of some observed Galactic GCs. However, the sampling of initial cluster properties does not aim to reproduce the full diversity of star clusters found in the local Universe or at higher redshifts. Important parameters such as the initial mass function, binary properties, and the distribution of metallicities, while varied in our models, do not encompass all possibilities observed in nature. Although we include several different metallicities, the set remains limited, and metallicity plays a crucial role in the evolution of massive stars and the formation of BHs. Consequently, our results may not capture the full impact of environmental variation on sBBH formation and ejection.

Our modeling of BH and sBBH production employs prescriptions for stellar winds, supernova kicks, and common-envelope evolution that are subject to uncertainties and ongoing revision. The formation of BHs via fallback and the treatment of pair-instability supernovae, as well as the assignment of natal spins and GW recoil kicks, all depend on parameters that remain uncertain. These choices can affect the numbers, masses, and orbital properties of escaping BBHs. In particular, the retention and formation of second-generation BHs (via hierarchical mergers) are sensitive to the adopted prescriptions for spin magnitude, spin orientation, and recoil velocity.

Another important caveat concerns the detection predictions. Our calculations of signal-to-noise ratios (S/N) and parameter estimation precisions are based on the Fisher information matrix (FIM) method and on idealized assumptions about detector sensitivity and observing conditions. In this work, we adopt $\varrho=8$ as the S/N threshold for successful detection in both low- and middle-frequency observations, and set $\varrho=5$ for each frequency band when identifying multiband sBBH sources. However, higher S/N thresholds may be required in practice; for example, \citet{Moore2019} suggest that $\varrho=15$ is a more appropriate threshold for space-based missions, given the large number of template banks needed in real data analysis. The estimated detection rates in this paper may decrease significantly if a higher S/N threshold is applied. In addition, we assume a continuous 4-year observation period starting simultaneously for LISA, Taiji, and AMIGO, without accounting for potential gaps or differences in real observation schedules. In practice, some sources may only be detectable in certain frequency bands if detector observing windows do not overlap.

The FIM method, while computationally efficient, is most accurate at high S/N and can overestimate parameter estimation precision by up to an order of magnitude compared to full Bayesian analyses \citep[e.g.,][]{2013PhRvD..88h4013R,2022arXiv220702771I,2014ApJ...789L...5K,2014PhRvD..89d2004G,2015PhRvD..91d2003V,2021Holgado,2024Xuan}. The results are sensitive to the specific parameters and waveform models chosen. In future work, we will adopt full Bayesian analyses to provide more robust parameter estimation.

Finally, our current analysis focuses exclusively on sBBHs that are ejected from GCs and subsequently merge in isolation. Merging binaries that are retained within clusters and experience further dynamical interactions are not included here. We also note that PN corrections are not included in the few-body integrations, similar to the treatment in \citet{kremer2018} so highly eccentric in-cluster mergers driven by GW dissipation during close encounters are not captured in the present analysis. Such in-cluster mergers may have significantly higher eccentricities and different mass distributions than escaping binaries, potentially making them promising multiband GW sources as well\citep{kremer2019}. We will include these populations in future studies to further investigate the prospects for distinguishing sBBH formation channels via multiband GW observations.

Taken together, these caveats highlight the importance of continued efforts to improve cluster modeling, incorporate a wider range of physical effects and initial conditions, and compare theoretical predictions with observations as the next generation of gravitational wave detectors comes online.

\section{Data Availability} 
{The underlying data for the $13409$ escaping BBHs that will merge within a Hubble time, obtained from the 268 GC models simulated with the MOCCA code are publicly available at \href{https://doi.org/10.5281/zenodo.15858711}{https://doi.org/10.5281/zenodo.15858711}}.

\section{Acknowledgments}
\begin{acknowledgments}
This work is partly supported by the National Key Program for Science and Technology Research and Development (Grant No. 2020YFC2201400), the National Natural Science Foundation of China (Grant No. 12273050, 11991052), the Strategic Priority Program of the Chinese Academy of Sciences (Grant No. XDB XDB0550300), and the Postdoctoral Fellowship Program of CPSF (GZC20240127). AA and SA acknowledge support for this paper from project No. 2021/43/P/ST9/03167 co-funded by the Polish National Science Center (NCN) and the European Union Framework Programme for Research and Innovation Horizon 2020 under the Marie Skłodowska-Curie grant agreement No. 945339. For the purpose of Open Access, the authors have applied for a CC-BY public copyright license to any Author Accepted Manuscript (AAM) version arising from this submission. AH, MG, GW, LH were supported by the NCN through the grant 2021/41/B/ST9/01191.
\end{acknowledgments}

%



\software{astropy \citep{2013A&A...558A..33A},  
          pycbc \citep{2019PASP..131b4503B}, 
          }




\bibliography{example}{}
\bibliographystyle{aasjournal}



\end{document}